\documentclass[twocolumn,pra]{revtex4-1}
\usepackage[T1]{fontenc}
\usepackage[utf8]{inputenc}
\usepackage{amsmath,amssymb,bbold}
\usepackage{color}
\usepackage{graphicx}
\usepackage{amssymb}
\begin{document}
\title{Band Structure and Topological Properties of Graphene in a Superlattice Spin Exchange Field.}
\author{ Luis Brey}
\email{Electronic address: brey@icmm.csic.es}
\affiliation{Instituto de Ciencia de Materiales de Madrid, CSIC, 28049 Cantoblanco, Spain}
\author{ A.R.Carvalho}
\affiliation{Instituto de F\'{\i}sica, Universidade Federal Fluminense, 24210-346 Niter\'oi, RJ, Brazil}
\author{H.A.Fertig}
\affiliation{Department of Physics, Indiana University, Bloomington, Indiana 47405, USA}
\date{\today}
\pacs{}

\begin{abstract}
We analyze the energy spectrum of graphene in the presence of spin-orbit coupling
and a unidirectionally periodic Zeeman field, focusing on the 
stability and location of Dirac
points it may support.  It is found that the Dirac points at the $K$ and $K'$
points are generically moved to other locations in the Brillouin zone, but that
they remain present when the Zeeman field $\vec{\Delta}(x)$ integrates to zero
within a unit cell.  A large variety of locations for the Dirac points is
shown to be possible: when $\vec\Delta \parallel \hat{z}$ they are shifted
from their original locations along the direction perpendicular to the superlattice
axis, while realizations of $\vec\Delta(x)$ that rotate periodically move
the Dirac points to locations that can reflect the orbit of the rotating electron 
spin as it moves through a unit cell.  When a uniform Zeeman field is applied
in addition to a periodic $\vec\Delta \parallel \hat{z}$ integrating to zero,
the system can be brought into a metallic, Dirac semimetal, or insulating state,
depending on the direction of the uniform field.  The latter is shown to be
an anomalous quantum Hall insulator.
\end{abstract}

\maketitle
\section{Introduction}
Many of the the proposed technological applications of graphene rely on the possibility of opening gaps at the Dirac points in its band structure \cite{Avouris_2007,Novoselov_2012}.
Such gaps
would also allow the experimental study of remarkable exotic physical effects that have
been predicted by theory \cite{Guinea_2009,Katsnelson-book}.
The interest in this system increases greatly
if in addition to the band gap, the resulting band structure supports non-trivial topology \cite{Hasan-2010,Qi-2011}.
Because of the high mobility of its carriers, graphene potentially could be an ideal system for studying  electronic properties of a system with momentum-space
Berry's curvature \cite{Xiao_2010}.

Different strategies have been proposed to endow
graphene with topological character.
One approach involves
the commensurate  stacking of graphene on
hexagonal boron nitride \cite{Gorbachev_2014,Pablo_2014a,Song_2015}.
Another strategy is to induce spin-orbit coupling in the system and thereby
open gaps in the spectrum. Along these lines, it has been proposed that a
combination of an exchange field and Rashba spin-orbit coupling (SOC)
will open such gaps at the Dirac points in monolayer
\cite{Qiao_2010,Qiao_2012,Zhang_2012} and bilayer
\cite{Tse_2011,Qiao_2013} graphene.
This gapped phase is  a realization of the quantum anomalous  Hall (QAH) insulator in graphene, with a non-vanishing Chern number and accompanying gapless edge states \cite{Qiao_2010}.
The two ingredients for the QAH phase can be provided by Fe adatoms \cite{Qiao_2010} or by adsorption of randomly distributed heavy adatoms in the presence of a proximity-induced exchange field \cite{Brey_2015}. Coupling between graphene and an
antiferromagentic insulator also could induce the anomalous  
quantum  Hall phase \cite{Qiao_2014}.

In this work we discuss the effects of a unidirectional superlattice Zeeman field on the electronic and topological properties of graphene.  This could be induced by spin-exchange
with an appropriate substrate, an external magnetic field, or some combination of the two.
When spin is a good quantum number, this is equivalent for each spin species to
a unidirectional periodic electrostatic potential, which can induce new Dirac points in graphene \cite{Brey_2009,Park_2009,Barbier_2010, Arovas_2010,Burset_2011}.
In what follows we investigate how this behavior is modified by Rashba spin-orbit coupling (SOC),
quantified by an energy scale $\lambda^R$.  Our analysis, which works in the continuum
limit and ignores
intervalley scattering, yields results which are essentially the same for
states near the $K$ and $K'$ points, so we will explicitly discuss results
for the first of these. We have  checked with microscopic tight-binding calculations the accuracy  of this approximation.

In the absence of SOC, if the exchange potential
induces an effective Zeeman field purely in the $\pm\hat{z}$ direction, $\Delta_z(x)$,
for each spin there is an effective
periodic potential, which throughout this paper we assume varies in the $\hat{x}$
direction. We mostly focus on situations in which the net exchange field,
$\int dx\Delta_z(x) = 0$. For weak and/or short wavelength potentials,
the main effect of this is to
render the dispersion around the Dirac points associated with each spin anisotropic
\cite{Park_2008}.  We find that SOC splits these spin degenerate Dirac cones, pushing
their zero energy points away from the $K$ point, onto the $k_y$ axis.
For stronger or longer wavelength $\Delta_z(x)$, at  zero Rashba SOC, new
Dirac points are already present on the
$k_y$ axis \cite{Brey_2009,Park_2009}. We find that
with increasing $\lambda^R$ these higher order Dirac points become gapped,
except for narrow ranges of $\lambda^R$ where the different species of Dirac points may
be made to approach one another.
For most values of $\lambda^R$, however, only two
Dirac points remain, residing symmetrically along the $k_y$ axis on either
side of the $K$ point.  In these situations, an application of a uniform
Zeeman field, or an imbalance in the positive and negative regions of
$\Delta_z$ induces gaps in these last two Dirac points, and the band
becomes topological, supporting an QAH effect.

We also consider the effects of in-plane Zeeman fields $\Delta_x$ and $\Delta_y$.
These can arise if an in-plane magnetic field is applied in addition to exchange
coupling to a magnetized system.
They can also be present if the magnetized
substrate contains a one-dimensional array of oppositely oriented
domains.  Regions between these regions are domain walls, and these generically
contain in-plane fields in the $\hat x$ direction (Bloch walls) or $\hat y$ direction
(N\'eel walls).  We shall see that the effect on the energy level structure
depends on which of these two cases is realized, and also on whether
neighboring domain walls are aligned or anti-aligned.
In some cases there is no qualitative effect on the energy dispersion, in others
the Dirac points may by moved into the $k_x-k_y$ plane, and in still others
they may reside on the $\vec{k}$ axis along the superlattice direction rather than
perpendicular to it.
An interesting aspect of
in-plane Zeeman fields induced by domain walls
is that the locations of their zero energy states reflect the
closed loop an electron spin traces on the Bloch sphere when it
traverses a single period of
the superlattice.

Because the carbon atoms are very light, in pristine graphene the SOC is extremely small \cite{Huertas-2006,Min-2006,Chico-2009}. Recent studies have shown that Rashba SOC is enhanced when graphene is bent with respect to its planar geometry \cite{CastroNeto-2009,Balakrishnan-2013}.
A combination of this enhanced SOC and a modulated exchange field could allow the realization
of configurations presented in this work. Several other approaches could produce a modulation of the effective Zeeman field.  
The first involves ``Origami structures'' \cite{Costa_2013}, in which
graphene sheets are periodically folded so that the effective magnetic field
oscillates in sign as one moves
through the graphene layer.  A second possibility is to consider
carbon nanotubes in contact with an insulating ferromagnet \cite{Cespedes_2004}.
In these cases, curvature in the graphene sheet enhances $\lambda ^R$.
In a third approach, one may consider a graphene layer placed in
close proximity to a ferromagnet
with oppositely oriented magnetic domains alternating along some direction.
Finally, an exchange field coupling to the electron spin may be created by depositing hydrogen atoms on the graphene surface \cite{Palacios_2008,Palacios_2010,McCreary_2012}.

This article is generally divided into two parts.  In the first we describe our analysis
of the system when the exchange field is purely along the $\hat{z}$ direction.
We present a perturbative analysis of the system, as well as tight-binding study results.
Finally we present a non-perturbative analysis (with details in an appendix)
that both confirms the perturbative results, and shows what happens to the higher order
Dirac points present in the absence of SOC.  In the second part, we discuss the effects
of additional field components in the $\hat{x}-\hat{y}$ plane.  We begin with a
perturbative analysis of uniform in-plane fields.  We then present analyses of two
models that include domain walls, one using a uniformly rotating exchange field, and
the other a piecewise-constant exchange field. This is followed by a discussion of
numerical tight-binding studies.  We finally conclude with a discussion and summary.

\section{Unidirectional Zeeman Fields}
We begin by considering models in which the effective Zeeman field is always aligned in the
$\hat{z}$ direction.
The Hamiltonian for such systems consist of three terms,
\begin{equation}
H=H_0+H_R+H_{\Delta},
\label{Htotal}
\end{equation}
with $H_0$ the kinetic energy, $H_R$ the Rashba SOC, and $H_{\Delta}$ is a spin-orienting
term.  In the absence of SOC,  the conduction and valence bands touch
to form Dirac points at the
${K}$ and ${K}'$ points in the Brillouin zone.
Near these points the low energy physics is described by the Hamiltonian
\begin{equation}
H_0=\hbar v_F ( - i s \sigma _0  \! \otimes  \!  \tau _x  \,  \partial _x  -i  \sigma _0 \! \otimes  \! \tau _ y  \, \partial _y) \,,
\label{Dirac}
\end{equation}
where $v_F$ is the Fermi velocity of the electrons,
$s=1$ ($s=-1$) for the $K$ ($K'$) valley, and  $\vec\tau $ is
a vector of Pauli matrices acting on a spinor specifying the amplitudes of the
wavefunction on the $A$ and $B$ sublattices of graphene.
A second vector of Pauli matrices $\vec\sigma$ acts on the real spin
degree of freedom of an electron, and accompanying both the $\vec\tau$ and
$\vec\sigma$ matrices are corresponding unit matrices, $\tau_0$ and $\sigma_0$.

Rashba SOC appears because of broken mirror symmetry, by interaction with
a substrate or induced by heavy adatoms \cite{Brey_2015}. In the
continuum approximation the SOC Rashba term takes the form \cite{Kane-2005},
\begin{equation}
H_{SO} ^R  =  \frac 1 2 \lambda  ^R ( -\sigma _x \otimes \tau _y +s \, \sigma _y \otimes \tau _x) \, .
\label{SOC}
\end{equation}
Finally we include a unidirectional superlattice effective Zeeman field,
which contributes to the Hamiltonian via
\begin{equation}
H_{\Delta}= \Delta_z (x) \sigma _z \otimes \tau _0.
\label{exchange}
\end{equation}
For this section we consider an antiperiodic exchange term, i.e.,
$\Delta_z (x+ L/2)$=$-\Delta_z (x)$, where $L$ is the period of the Zeeman field.

As noted in the Introduction, the analytic Hamiltonians we consider in this work
are continuum approximations which are appropriate when intervalley scattering
is negligible.  In general this occurs when the {Zeeman field} varies over
length scales which are large compared to the graphene lattice parameter.
Adopting this assumption, in the rest of this paper we will explicitly discuss
results only for the ${K}$ valley ($s=1)$.  Within this assumption, results for
the $K'$ valley are essentially identical.

\subsection{Symmetry Considerations}
\label{sec:symmetry}
In what follows we will be mostly interested in the energy spectrum of this system
as a function of $\vec{k}$, the deviation of the momentum from the Dirac point.
It is helpful to begin by considering some symmetries of the system.
One of these is a chiral symmetry operation \cite{Sun_2010} $T$, consisting
of a combined operation of a sublattice chiral operator,
$\sigma _0 \otimes \tau _z$, and a shift operator $x \rightarrow x+L/2$, so that
\begin{equation}
T^{\dagger} H(x) T = (\sigma _0 \otimes \tau_z ) H(x+ \frac L 2) (\sigma _0 \otimes \tau _z) = -H(x).
\end{equation}
Because $T$ commutes with the translation operator, it preserves wavevector, so the
anticommutation property of $T$ with $H$ guarantees that for any state at
wavevector ${\vec k}$ of energy $\epsilon({\vec k})$, there is a corresponding
state at the same wavevector with energy $-\epsilon({\vec k})$.

A second observation is that, because
the Hamiltonian does not depend on  $y$, the momentum in the $y$ direction,
$k_y$, is a good quantum number. Writing a wavefunction in the form
$\vec\Phi_{k_x,k_y}(x,y) = e^{ik_yy}\vec\Psi(x)$, the effective
Hamiltonian acting on $\vec\Psi$ becomes $H(k_y) \equiv e^{- ik_yy} H e^{ik_yy}$,
and has the property
\begin{equation}
H(-k_y) =  \sigma _z  \! \otimes  \! \tau_0 \,  H ^* (k_y) \,  \sigma _z \! \otimes  \! \tau _0.
\end{equation}
This implies that for any zero mode of Eq. \ref{Htotal} appearing at a particular
value of $k_y \equiv k_y^*$, we can construct another zero mode with $k_y=-k_y^*$.


Finally,  the Hamiltonian commutes with a generalized mirror operator
$X$=$\sigma _x \otimes \tau _ y I _x$,
where $I_x f(x)=f(-x)$, and assuming $\Delta (-x)=-\Delta (x)$, one may easily show $[X,H]=0$.
Moreover,
$X$ satisfies  $\{T,X \}$=0 and  $X^2=1$. This means that states which are chiral
partners (in general with energies $E$ and $-E$) have opposite eigenvalues of the
operator $X$.  This allows the system to support Dirac points; i.e., when
$\epsilon({\vec k})$ approaches 0 as ${\vec k}$ is varied, it is not repelled by
its chiral partner with which it becomes degenerate.  The $X$ operator can be
exploited to help locate zero modes for this system, as we show below.


\subsection{Perturbation Theory}
\label{sec:pert_theory}
When the spin-orbit coupling is zero ($\lambda ^R$=0),
the Hamiltonian consist of two uncoupled blocks
with well-defined spin quantum numbers,
each of which supports a zero energy Dirac point at ${\vec k}=0$ \cite{Brey_2009,note1}.
For small $\lambda^R$ one can obtain the effective Hamiltonian in the vicinity of this
point using degenerate perturbation theory; i.e., by projecting the Hamiltonian onto
the zero energy basis for $\lambda^R=0$.  The four zero energy states have the explicit forms
\begin{widetext}
\begin{equation}
\psi_{\uparrow,+} \! =\frac 1 {\sqrt{L}}  \left ( \begin{array}{c} \cos \theta (x) \\ -i \sin \theta (x) \\ 0 \\ 0 \end{array} \right ) , \,
\psi_{\uparrow,-}\!=\frac 1 {\sqrt{L}}  \left ( \begin{array}{c} -i \sin \theta (x) \\ \cos\theta (x) \\ 0 \\ 0 \end{array} \right ) , \,
\psi_{\downarrow,+}\!\ = \!\frac 1 {\sqrt{L}} \left ( \begin{array}{c} 0 \\ 0 \\ \cos \theta (x) \\ i \sin \theta (x)  \end{array} \right ) , \,
\psi_{\downarrow,-}\!=\!\frac 1 {\sqrt{L}}  \left ( \begin{array}{c}  0 \\ 0  \\ i \sin \theta (x) \\ \cos\theta (x) \end{array} \right ) ,
\label{basis}
\end{equation}
with
\begin{equation}
\theta (x)= - \frac 1 {\hbar v_F} \int _0 ^x \Delta (x) dx.
\end{equation}
Multiplying the states in Eq. \ref{basis} by $e^{i\vec k \cdot \vec r}$, the projected
Hamiltonian (Eq.\ref{Htotal}) in the resulting basis takes the form
\begin{equation}
H= \left ( \begin{array}{cccc}
0 & \hbar v_F (k_x -i k_y f_0) & 0 & -i \frac {\lambda ^R} 2 ( 1+ f_0 ) \\
\hbar v_F (k_x +i k_y f_0) & 0 &  i\frac {\lambda ^R} 2 ( 1- f_0 ) & 0 \\
0&  -i\frac {\lambda ^R} 2 ( 1- f_0) & 0 & \hbar v_F (k_x -i k_y f_0 )      \\
 i \frac {\lambda ^R} 2 ( 1+ f_0  )& 0 & \hbar v_F (k_x +i k_y f_0 ) &0
\end{array} \right ),
\label{Heff}
\end{equation}
\end{widetext}
with
\begin{equation}
f_0 = \frac 1 L \int _0 ^L  \cos {2 \theta (x)} \, .
\end{equation}
The effective Hamiltonian Eq. \ref{Heff} has zero energy solutions. These zero modes appear at momentum ${\vec k}^*$, with
\begin{equation}
k_x ^* =0 \, \, , \, \, k_y^* = \pm \frac {\lambda ^R}{2 \hbar v_F f_0}\sqrt{1- f_0 ^2}.
\label{newD}
\end{equation}
One may further expand Eq. \ref{Heff} around ${\vec k}^*$ by projecting it onto the two zero
mode states at that point.  Diagonalizing the resulting $2 \times 2$ Hamiltonian
yields the dispersion law
\begin{equation}
\epsilon (\tilde k_x, \tilde k_y )= \hbar v_F \sqrt{1-f_0^2} \sqrt{ \tilde k_x ^2 + f_0^2 \tilde k_y ^2},
\end{equation}
where $\tilde {k}_{x,y}={k_{x,y}} - {k}_{x,y}^*$ is the momentum measured relatively to the Dirac points, Eq. \ref{newD}.


\subsection{Tight Binding Studies}
The band structure of pristine graphene is well-described by a tight-binding model with hopping $t \sim 2.7$eV between nearest neighbor atoms on a honeycomb lattice, represented by
\begin{equation}
H_0=-t \sum _{<i,j>,\sigma}  c^{\dag} _{i,\sigma} c _{j,\sigma} 
\label{TB-Graphene}
\end{equation}
where $c^{\dag}_{i,\sigma} $ creates  an electron at site $i$ with spin $\sigma$.
Rashba SOC in this system can be implemented as a spin-dependent hopping
between nearest neighbors
of the form
\begin{equation}
H_R \! =  \! i \frac {\lambda ^{R}} 3 \!\!\!\!  \sum _{<i,j>} \sum_{s ,s'}   {\hat z}\cdot ( \vec \sigma \!
  \times \! \hat { u}_{i j} )_{s,s'}  \,  c^{\dag} _{i,s} c _{j,s'},
\label{RSOtb}
\end{equation}
where $\hat { u}_  {ij} $ is a unit vector pointing from site $j$ to site $i$ and
 $\vec  \sigma $ is the vector of spin Pauli matrices.  Finally
the Zeeman coupling is implemented through a term of the form
 \begin{equation}
H_{\Delta} \! =    \sum _{i,s,s'}   \Delta (i)    c^{\dag} _{i,s }
(\sigma _z)_{s,s'} c _{i,s'}
\label{ExchangeTB}
\end{equation}
where  $\Delta (i)$ is the  periodic Zeeman field.

For simplicity we consider here
a Kronig-Penney model Zeeman field.  Writing this
in the form $\Delta (x)$= $\Delta _0 \, \rm {sgn}(x)$ for $|x|<L/2$, one finds for the
expressions in the last subsection
\cite{Arovas_2010,Barbier_2010},
\begin{equation}
f_0 = \frac {|\sin u|} {u} \, \, \, \, {\rm with} \, \, \, \,  u=\frac {\Delta _0 L}{2 \hbar v_F} .
\label{f0}
\end{equation}

\begin{figure}[htbp]
\includegraphics[width=\columnwidth,clip]{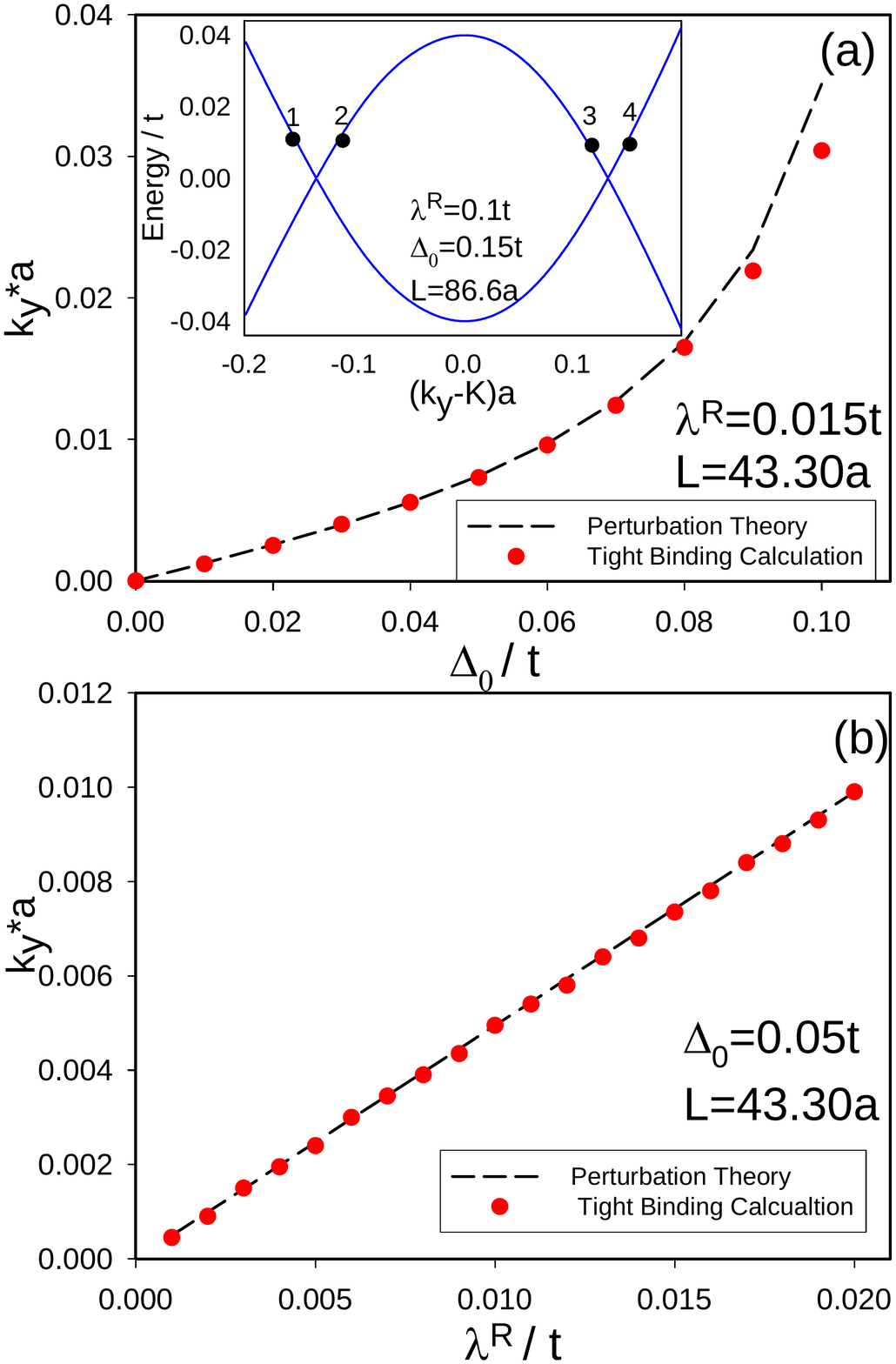}
\caption{(Color online) (a) and (b) position of the  split Dirac points as a function of the intensity of the periodic exchange field  and Rashba SOC respectively. Red dots indicate the tight binding results, whereas the lines are continuum results
obtained from Eq. \ref{newD} and  Eq. \ref{f0}. In the inset of (a) we plot a typical tight binding band structure near the original Dirac point. The quantity $a$ is the graphene lattice parameter.}
\label{Bandas}
\end{figure}

A typical band dispersion obtained from diagonalizing the tight-binding Hamiltonian
is illustrated in the inset of Fig. \ref{Bandas}(a). For $\lambda^R=0$ one obtains two
degenerate Dirac cones centered at the $K$ point.  As expected from our perturbative
analysis, for $\lambda^R \ne 0$ these are repelled down the $k_y$ axis, settling at values
$k_y=\pm k_y^*$ that depend on $\lambda^R$, $L$, and $\Delta_0$.
In Fig. \ref{Bandas}(a) and (b) we plot the dependence of $k_y^*$ on the Rashba and
exchange field respectively, as obtained from the tight-binding
calculation and from Eqs. \ref{newD} and \ref{f0}.
The agreement between the tight-binding calculations and our perturbative
analysis is very good when the parameter $u=\frac {\Delta _0 L}{2 \hbar v_F}$ is smaller than $\pi$. For larger values, new Dirac points emerge in the spectrum of the decoupled spin Hamiltonians \cite{Brey_2009,note1}.  The projected Hamiltonian (in the states described
by Eq. \ref{basis})
is not sufficient to capture this physics.  Not surprisingly, as this regime is approached
the predictions of Section \ref{sec:pert_theory} become inaccurate.  Below we discuss the
behavior of the higher-order Dirac points associated with large $u$ when $\lambda^R \ne 0$.

\subsection{Edge State Interpretation, Reflection Symmetry, and Quantized Transport}

\begin{figure}[htbp]
\includegraphics[width=\columnwidth,clip]{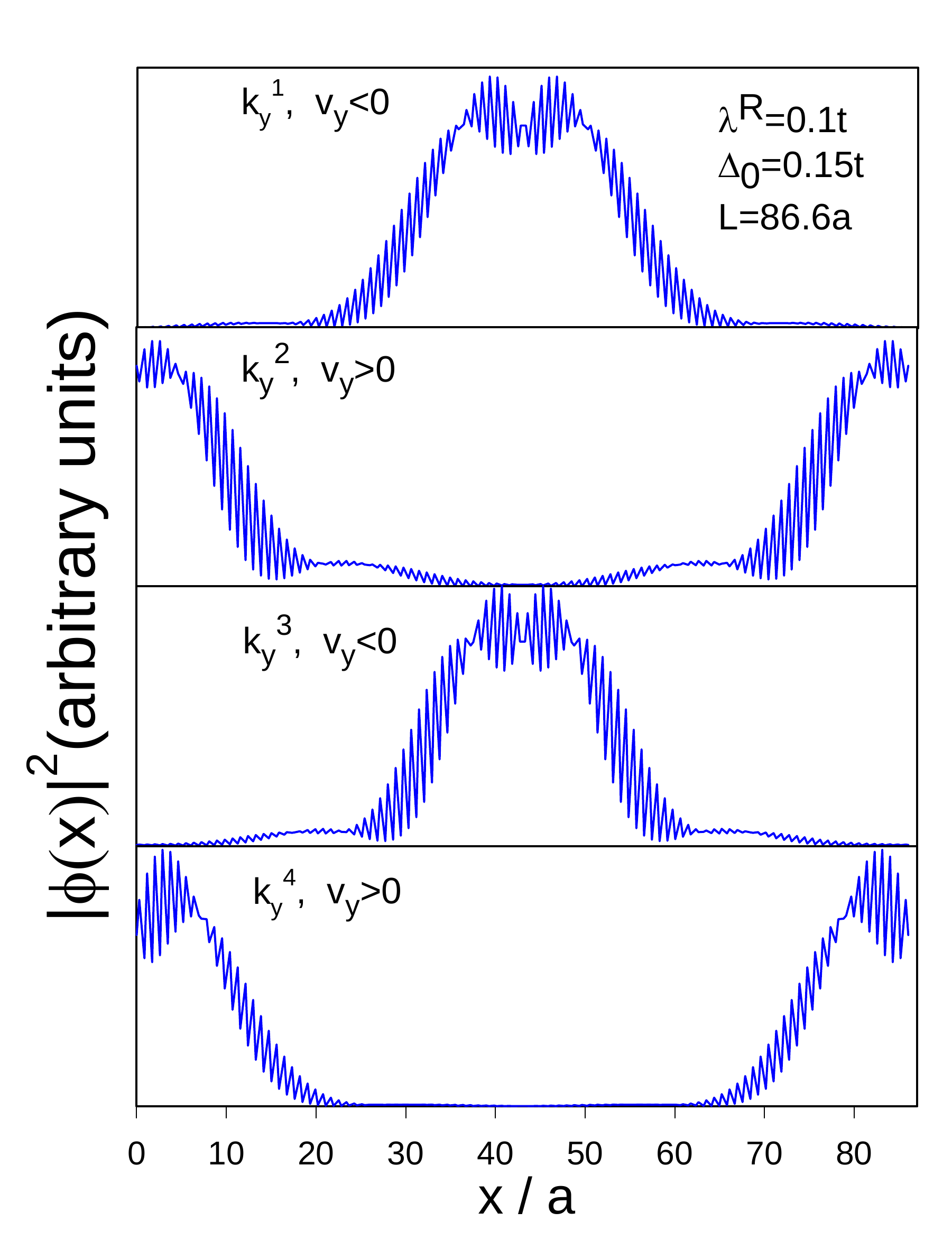}
\caption{(Color online) Square of the wave function as function of the position, of the four states, 1-4, marked in the inset of Fig.\ref{Bandas}(a), $v_y$ indicates the velocity of the states along the $\hat y$-direction. $a$ is the graphene lattice parameter.}
\label{wf}
\end{figure}

Some insight about the nature of the SOC-split Dirac points is obtained by examining wave functions of states near zero energy. In Fig. \ref{wf} we plot the absolute square of the wave functions corresponding to the four states marked in the inset of Fig. \ref{Bandas}.
Interestingly,
the states are largely localized at the (two) interfaces of the unit cell separating
regions of opposite signed Zeeman fields. Moreover, the direction of the
group velocity associated with
these states correlates with the sign of the Zeeman field steps around which they are located:
up-steps and down-steps support states of opposite velocity.  Notice that for a given
energy there are {\it two} such states at a given interface in the vicinity of the
$K$ point.  Two further states for each interface can be found near the $K'$ point.

These interface states are similar to those appearing in bilayer graphene in an electrostatic lateral confinement step \cite{Martin_2008,Li_2011}, and in periodically modulated bilayer graphene \cite{Barbier_2009,Killi_2011,Tan_2011}.
The origin of theses interface states is topological. Graphene in the presence of Rashba SOC and constant exchange coupling $\Delta$ is a quantum anomalous  Hall system \cite{Qiao_2010}
with Hall conductivity
\begin{equation}
\sigma _{xy} = \frac {e^2} h C ,
\end{equation}
where $C$=$2 \,\rm{sgn} (\Delta)$ is the Chern number associated with the full band, and the factor of 2 enters due to the two Dirac points (valleys) in graphene. Because of the different
Chern numbers in each region, at their interface there should appear four parallel velocity edge channels.  For a generic superlattice of such interfaces, their coupling will generally
gap out the spectrum; however, in the presence of an appropriate
reflection symmetry (specifically the $X$ operator above), for certain choices
of $\vec{k} \equiv \vec{k}^*$ there are zero energy states \cite{Killi_2011}.
This is the origin of the two Dirac points per valley that appear at this superlattice.

The above discussion emphasizes the role of reflection symmetry in the form of the operator
$X$ in leading to a gapless spectrum.  By contrast if this symmetry is broken one might
expect a gap to open \cite{Killi_2011}.  Moreover, since the system is not time-reversal symmetric, non-zero
quantized Hall transport may result.  The simplest perturbation one can introduce that
does this is an overall magnetic field in the $\hat z$-direction coupling to the spins,
which may be a result of an imbalance between the two directions of the exchange field,
or simply from an applied magnetic field weak enough that the coupling to orbital motion
may be ignored.  To the Hamiltonian in Eq. \ref{Heff} this adds a term of the form
\begin{equation}
\Delta H= \left ( \begin{array}{cccc}
b_z& 0 & 0 &0 \\
0 & b_z & 0 & 0 \\
0& 0 & -b_z & 0      \\
0& 0 & 0 & -b_z
\end{array} \right ).
\end{equation}
One can show this indeed opens a gap by examining the Hamiltonian $H+\Delta H$
when it is projected into the subspace of zero energy states of $H$ (Eq. \ref{Heff}).
Defining Pauli matrices ${\mathrm \mu}$ for which $\mu_x |\psi_{\sigma,\pm}\rangle
= \pm |\psi_{\sigma,\pm}\rangle$, where $\sigma=\uparrow,\downarrow$ denotes spin,
and $|\psi_{\sigma,\pm}\rangle$ are the kets corresponding to Eqs. \ref{basis},
linear combinations may easily be constructed such that
$\mu_z|\mu_z=\pm,\sigma\rangle = \pm |\mu_z=\pm,\sigma\rangle$.  In terms
of these one finds, for example at ${\vec k}^* = (0,k_y^*)$, two zero modes
which can be labeled by $\mu_z$,
\begin{eqnarray}
|1\rangle_{k^*} &\equiv& {{\sqrt{1-f_0}} \over {\sqrt{2}}} |\mu_z=1,\uparrow\rangle
+{{\sqrt{1+f_0}} \over {\sqrt{2}}} |\mu_z=1,\downarrow\rangle, \nonumber \\
|-1\rangle_{k^*} &\equiv& {{\sqrt{1+f_0}} \over {\sqrt{2}}} |\mu_z=-1,\uparrow\rangle
+{{\sqrt{1-f_0}} \over {\sqrt{2}}} |\mu_z=-1,\downarrow\rangle. \nonumber \\
\label{kstar_basis}
\end{eqnarray}
Projection of $H+\Delta H$ into the $(|1\rangle_{k^*},|-1\rangle_{k^*}$ basis
yields an effective Hamiltonian $H_{k^*}$ given by
\begin{equation}
\label{Hkstar}
H_{k^*} = v_F\sqrt{1-f_0^2}
\left(
\begin{array}{c c}
b_z \frac{f_0}{\sqrt{1-f_0^2}} & \tilde k_x - if_0\tilde k_y \nonumber \\
k_x + if_0\tilde k_y & -b_z \frac{f_0}{\sqrt{1-f_0^2}} \\
\end{array}
\right),
\end{equation}
where $\vec{\tilde{k}}= {\vec k}-{\vec k}^*$.  The energy eigenvalues are
$$\varepsilon(\tilde k_x,\tilde k_y) = \pm \sqrt{v_f^2(1-f_0^2)
(\tilde k_x^2 +f_0 \tilde k_y^2) + (f_0 b_z)^2 },$$
which is in general gapped, as expected.  Clearly one may use the basis in
Eq. \ref{kstar_basis} to examine the effects of Zeeman fields in the $\hat x$
and $\hat y$ directions as well; however we defer this discussion to the
next section.

\begin{figure}[htbp]
\includegraphics[width=\columnwidth,clip]{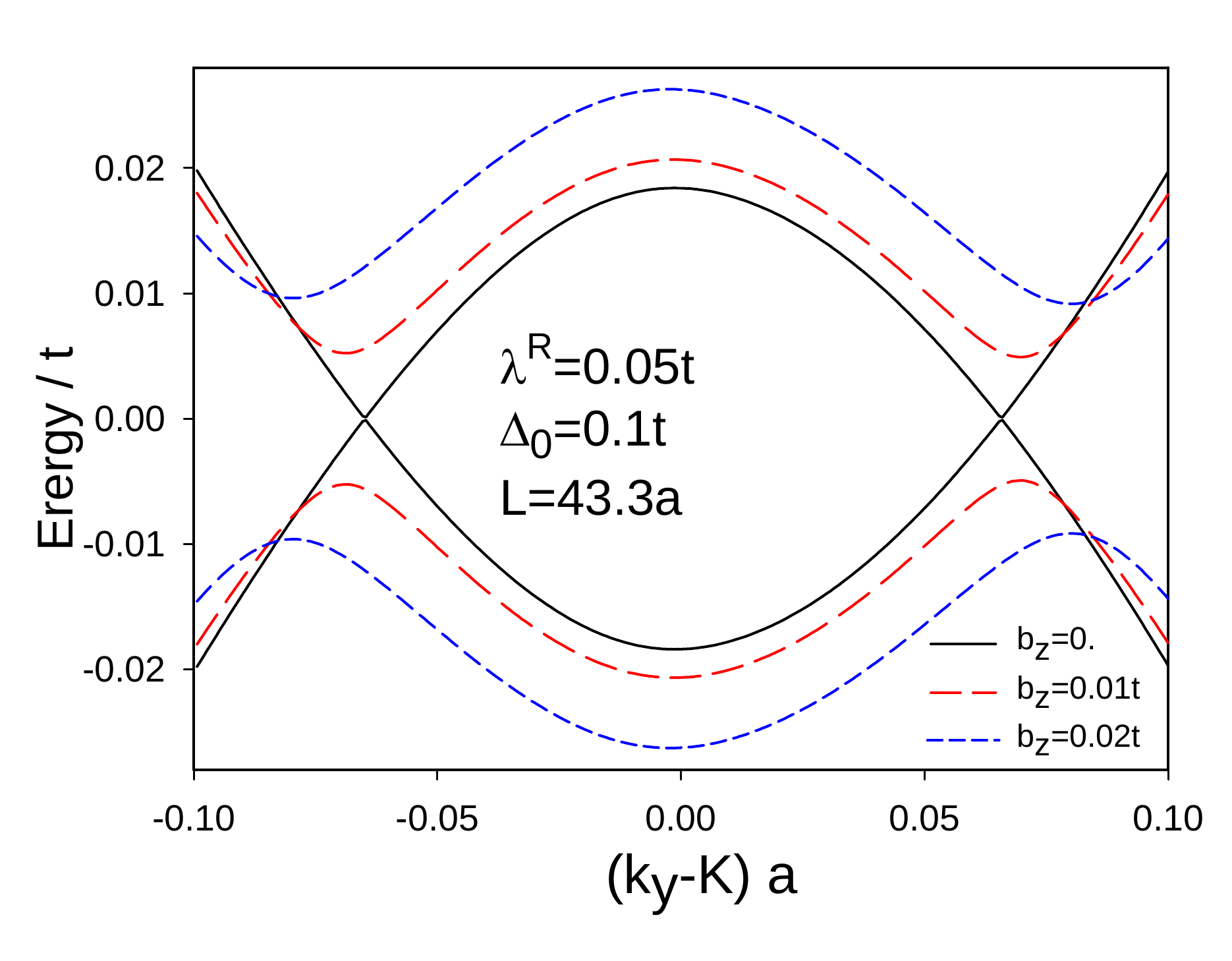}
\caption{(Color online) Tight-binding band structure near the Dirac point $K$ obtained  with the parameters
$\lambda ^R$=0.05$t$, $\Delta _0$=0.1$t$, $L$=43.3$a$ and different values of an overall exchange field pointing in the $z$-direction, $b_z$. }
\label{Deltac}
\end{figure}

\begin{figure}[htbp]
\includegraphics[width=\columnwidth,clip]{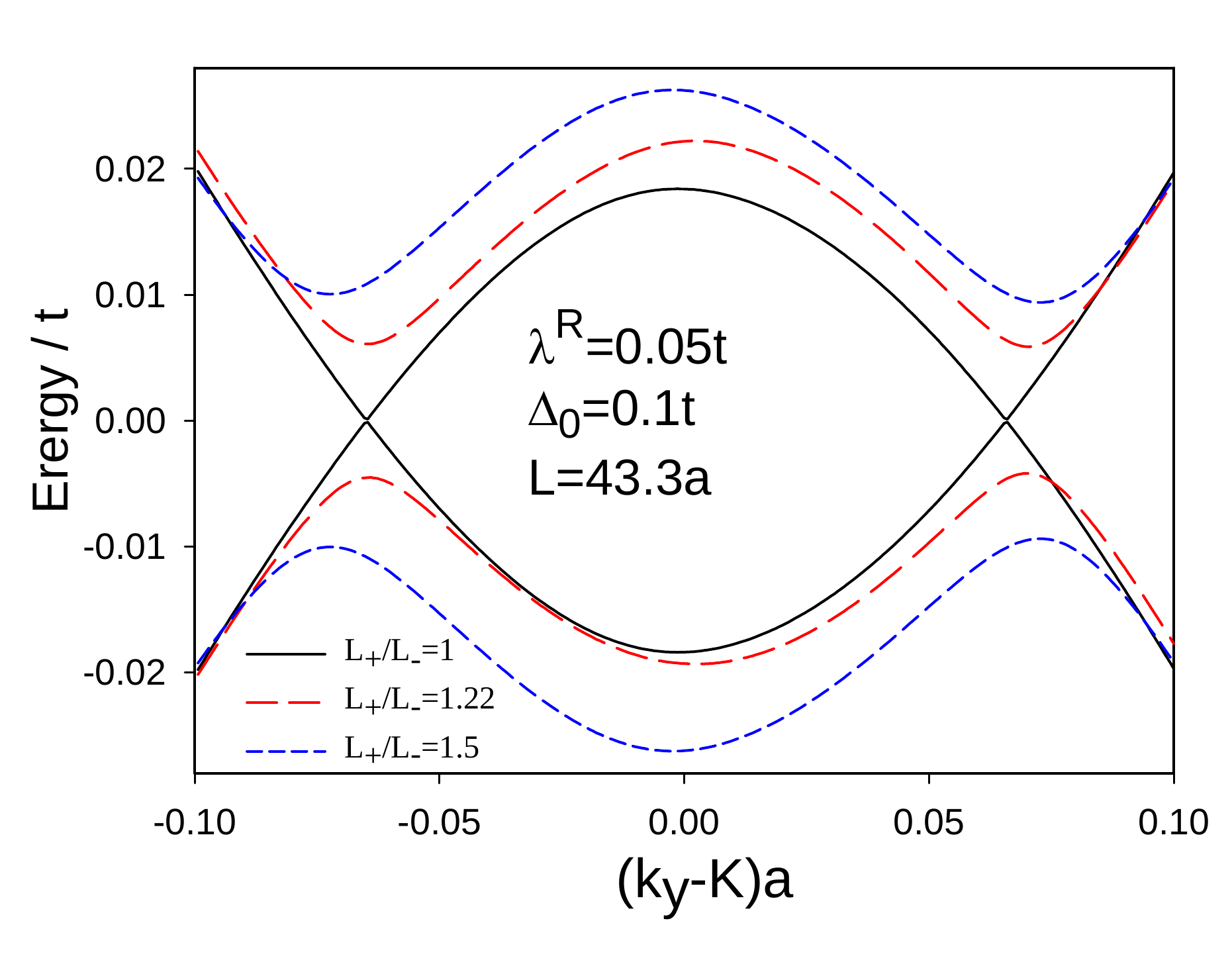}
\caption{(Color online) Tight-binding band structure near the Dirac point $K$ obtained  with the parameters
$\lambda ^R$=0.05$t$, $\Delta _0$=0.1$t$, $L$=43.3$a$ and different values of the ratio $L_+ /L_-$, see text.}
\label{Asim}
\end{figure}

Qualitatively, the gap opening is associated with a net effective Zeeman field
which is predominantly up or down in the $\hat{z}$ direction, so one may expect
that the Chern number associated with the resulting positive or negative energy
band closest to zero energy will be the same as for the corresponding system
with a uniform Zeeman field.  We have confirmed that this is indeed true.
In Fig. \ref{Deltac} we plot numerically obtained energy levels
near the $K$ point, in the presence of Rashba SOC, a piecewise-constant
periodic exchange field,
and a constant Zeeman field $\Delta _c$. For any finite $\Delta _c$ we find the
spectrum is gapped.
Similar physics applies
if the unit cell is asymmetric. Fig. \ref{Asim} illustrates numerical results for
energy levels in
such a case, where the periodic exchange field has larger regions of
positive exchange field than negative exchange field; i.e., $L_+/L_- > 1$.
Again the spectrum is gapped.

With such numerical tight-binding results we can compute
the Hall conductivity associated with a band, which is a direct measure of
its Chern number.  Specifically,
\begin{equation}
\sigma _{xy}  = -2 \frac {e ^2 \hbar} S  \sum _{n,n',{\bf k}} \frac {{\rm Im} \left ( <n {\bf k}|v _{x}  | n' {\bf k} >< n' {\bf k}|v _y |n {\bf k}>  \right ) } {(\varepsilon _{n, {\bf k} }- \varepsilon _{n',{\bf k}} ) ^2},
\end{equation}
where $S$ is the sample area and the velocity operator is given by
\begin{equation}
{\bf v} =- \frac i {\hbar} t \sum  _{<i,j>,\sigma} ({\bf r}_i-{\bf r}_j) c^{\dag} _{i,\sigma} c _{j,\sigma},
\end{equation}
with ${\bf r_i}$ the position of carbon atom at site $i$
For both cases described above one finds within numerical error that
\begin{equation}
\sigma _{xy}  = 2 \frac {e^2} h \rm{sgn}(\bar \Delta),
\end{equation}
where $\bar \Delta$ is the average value of the effective
Zeeman field, including
the contribution from $\Delta_c$.
Thus these systems support an quantum anomalous Hall effect, in which there is quantized
Hall transport even in the absence of an orbital magnetic field affecting
the electron dynamics.

\subsection{Zero Modes Beyond Perturbation Theory}
We finish this section with a non-perturbative analysis of zero modes.
To this point we have focused on zero modes which are located at the $K$ and $K'$
points for $\lambda^R=0$, and which split and move out the $k_y$ axis
when SOC is turned on.  In the absence of SOC, however, for sufficiently
large $|\Delta_z|L$ there will be further Dirac points for each spin
on the $k_y$ axis
\cite{Brey_2009,Park_2009}.  We would like to understand how these evolve
with $\lambda^R$.

\begin{widetext}
\begin{center}
\begin{figure}[htbp]
\includegraphics[width=\columnwidth,trim = 0 0 100 0,clip]{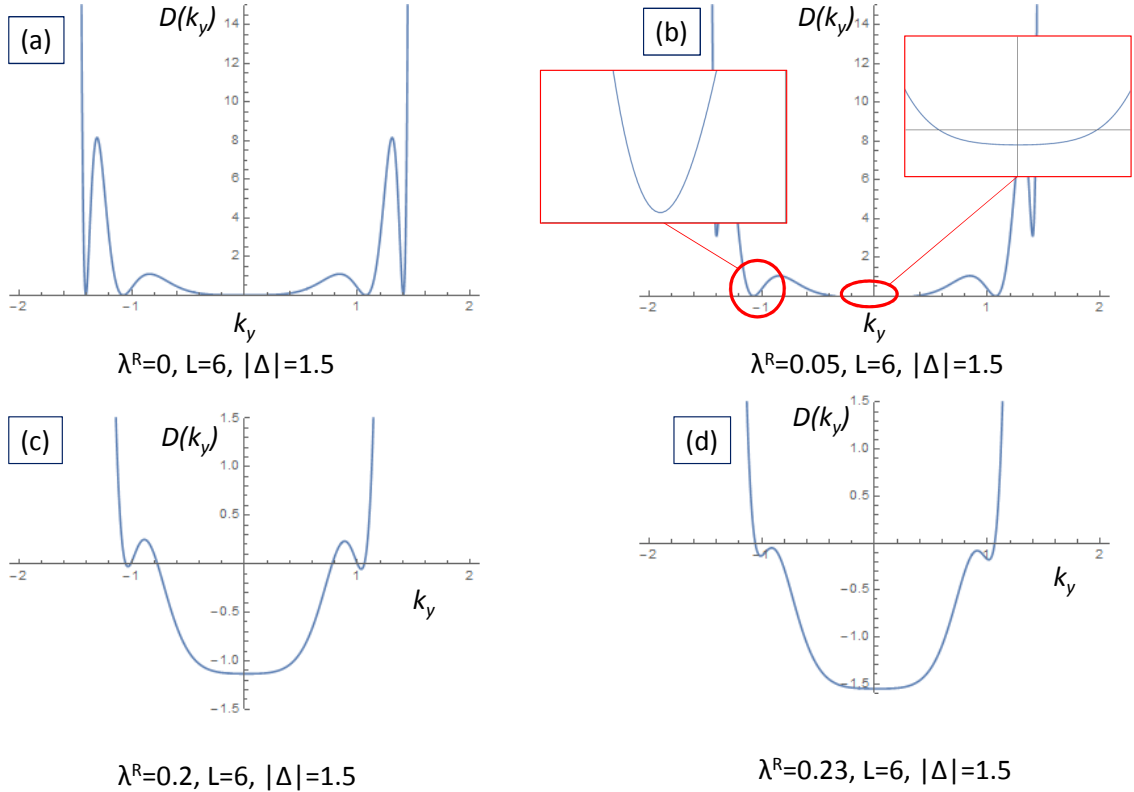}
\caption{(Color online) Plots of the function $D(k_y)$, which vanishes at values
of $k_y$ where there are zero energy states on the $k_y$ axis.  (a) $\lambda^R=0$,
illustrating Dirac points at $k_y=0$ as well as at non-vanishing values of $k_y$.
(b) $\lambda^R=0.05$.  Insets show that higher order vanishings of $D(k_y)$ are
lifted off the $k_y$ axis so that these Dirac points are eliminated, while the
degenerate Dirac points at $k_y=0$ are repelled down the $k_y$ axis but are
not eliminated.  (c) $\lambda^R=0.20$ and (d) $\lambda^R=0.23$ illustrate that multiple Dirac
points can emerge over narrow ranges of $\lambda^R$ at non-vanishing $k_y$.  For all
panels, unit cell length is $L=6a$ and $|\Delta|=1.5 \hbar v_F/a$.  All $\lambda^R$
values are in units of $\hbar v_F/a$. }
\label{nonpert}
\end{figure}
\end{center}
\end{widetext}

At present we are interested in the spectrum of Eq. \ref{Htotal} when
there is antisymmetry in the exchange field, $\Delta_z(-x)=-\Delta_z(x)$.
Eigenstates in such cases
may be characterized by a quantum number associated with
the reflection operator $X$.  Using a transfer matrix approach one may express
the condition that a zero mode exists for a given value of $k_y$ as being
met if the determinant a particular $2 \times 2$ real matrix $D(k_y)$
vanishes.
The demonstration of this is somewhat involved; details are provided
in the Appendix.  Fig. \ref{nonpert} illustrates a
typical evolution of $D(k_y)$ as $\lambda^R$
is raised from zero.  Exactly at $\lambda^R=0$ there are several zeros,
reflecting the multiple Dirac points expected for this situation \cite{Brey_2009}.
It should be noted there are {\it two} degenerate Dirac cones around each of
these points, one for each value of spin which is a good quantum number
in this situation.  When $\lambda^R$ is first raised from zero, these admix
and repel, leading to gaps everywhere except near $k_y=0$, where the degenerate Dirac
points at the origin are split into two and move down the $k_y$ axis.  With increasing
$\lambda^R$ these migrate further from the $K$ point, and may interact with residual
structure from the higher order Dirac points to produce extra Dirac points
over narrow ranges, as illustrated in Fig. \ref{nonpert}(c).  With the exception
of these situations, however, we find that only the two Dirac points predicted by
the perturbative analysis are
stable when SOC is present.  Finally, we note that this analysis verifies that
corrections at higher orders in perturbation theory do not open gaps at these
Dirac points.

\section{In-Plane Exchange and Zeeman Fields}

In this section we explore what happens to the spectrum when the effective Zeemen field
is not strictly oriented in the $\pm\hat{z}$ direction.  There are two reasons for considering
such situations.  Firstly, in addition to a periodic field induced by interaction
with a substrate, one may introduce
a uniform external magnetic field with an arbitrary size and orientation.
Such situations are interesting because, as we shall see, the spectrum is relatively 
sensitive to
these, so that one may in principle significantly modify the electronic structure in a
single system just by modifying the external field.  A second class of systems in which
in-plane Zeeman fields may be relevant are those in which electrons in the graphene sheet are
coupled to a system with magnetic domains with equal and opposite orientations.
Such magnetic domain structures are common in many ferromagnets, but the regions separating
domains are rarely sharp, and often support domain walls in which the magnetization
rotates continuously between opposing directions of the magnetization.  Thus the electrons
encounter localized regions of in-plane field.  Similarly, a periodic
Zeeman field can in principle
be implemented by subjecting a folded stack of graphene to an external magnetic field
(which we presume to be sufficiently weak that orbital effects of the field can be ignored.)
In this case the Zeeman field effectively rotates in the folded sections of the stack,
and once again localized regions of in-plane field will be present.

We begin by examining the simplest of these cases, a uniform Zeeman field present in addition
to the periodic $\Delta_z(x)$ field.

\subsection{Uniform In-Plane Zeeman Fields}
\label{sec:uniform_in_plane}

The impact of an externally imposed uniform Zeeman field on the
Dirac points near the $K$ point
can be easily assessed perturbatively, using the states appearing in Eq. \ref{kstar_basis}
in the same way as for $H_{k^*}$ (Eq. \ref{Hkstar}) to assess the effect of a uniform
field in the $\hat z$ direction.  Following the same procedure we find, for a perturbation
of the form $\Delta H_Z={\vec b} \cdot {\vec \sigma}$,
\begin{equation}
\label{Hkstarb}
H_{k^*}({\bf b}) = v_F\sqrt{1-f_0^2}
\left(
\begin{array}{c c}
b_x + b_z \frac{f_0}{\sqrt{1-f_0^2}} & \tilde k_x - if_0\tilde k_y \nonumber \\
k_x + if_0\tilde k_y & b_x-b_z \frac{f_0}{\sqrt{1-f_0^2}} \nonumber \\
\end{array}
\right).
\end{equation}
One can see that, as noted before, the $b_z$ contribution opens gaps at the Dirac points,
while $b_y$ has no qualitative effect on the spectrum, and $b_x$ shifts the Dirac point
away from zero energy.  For the last of these, the same analysis shows that $b_x$
component shifts the Dirac point at $k_y=-k^*$ in the opposite direction energetically,
so that $b_x$ by itself creates both electron-like and hole-like Fermi surfaces
when the graphene is undoped, enhancing the conductivity of the system.
The choice
of orientation of ${\vec b}$ in the $\hat{x}-\hat{z}$ plane thus allows one to tune the
system between an insulator and a metal.  Finally, at this level of approximation
it is interesting to note that $b_y$ has no effect on the spectrum.

\begin{figure}[htbp]
\includegraphics[width=\columnwidth,clip]{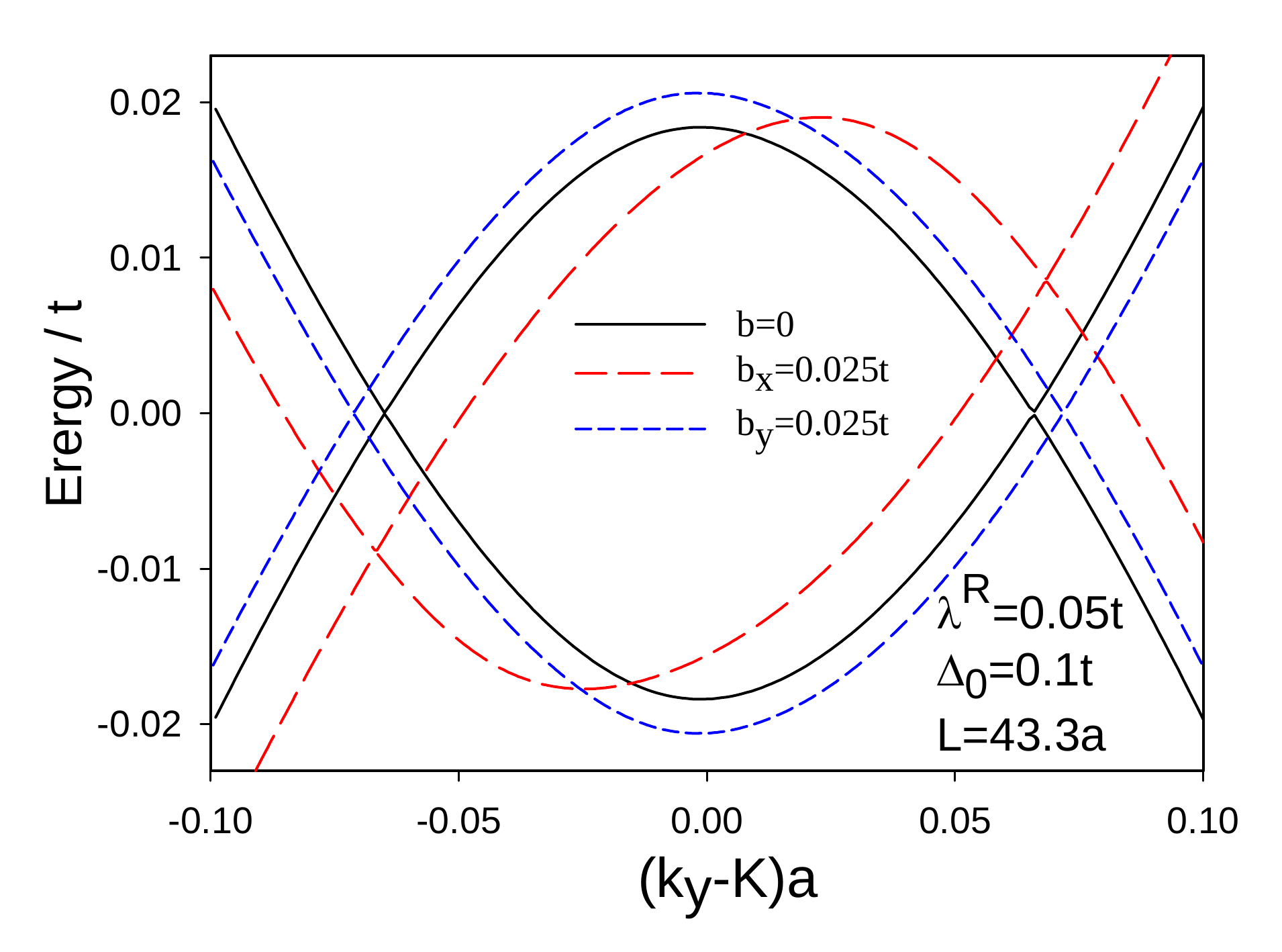}
\caption{(Color online) Tight-binding band structure near the Dirac point $K$ obtained  with the parameters
$\lambda ^R$=0.05$t$, $\Delta _0$=0.1$t$, $L$=43.3$a$ and   overall exchange fields pointing in the $x$ and $y$ directions. }
\label{Extra-B}
\end{figure}

Numerical tight-binding calculations support these results. Fig. \ref{Extra-B} illustrates
both the upward/downward energetic shifts of the Dirac
points with $b_x$ and the gap opening with $b_z$ occur as expected. Moreover, the stability
of the Dirac points at zero energy with respect to $b_y$ is confirmed, although they
move slightly in their location on the $k_y$ axis.  This behavior presumably occurs due
to corrections above linear order in $b_y$ which our perturbative analysis does not
capture.  We will see that analogous motion of zero modes is induced by domain wall
structures.

\subsection{Uniformly Rotating Zeeman Field}
\label{sec:uniform_rot}

As a simple model of a Zeeman field that rotates into different directions,
we consider one that rotates uniformly around either the $\hat{y}$ or
the $\hat{x}$ directions.  Explicitly,
\begin{eqnarray}
\Delta _z  & = & \Delta _0 \sin {\alpha (x)}, \nonumber \\
\Delta _x  & = & \Delta _0 \cos {\alpha (x)} \cos{\chi}, \nonumber \\
\Delta _y  & = & \Delta _0 \cos {\alpha (x)} \sin{\chi},
\label{rotation}
\end{eqnarray}
with $\alpha (x)$=$Gx$.  To analyze this model we form a basis by diagonalizing the
Hamiltonian $H_0+\Delta_z(x)\sigma_z$ precisely as in Section \ref{sec:pert_theory},
yielding four zero energy states at the $K$ point.  Expanding the full Hamiltonian
$H_0+\vec\Delta(x)\cdot \vec{\sigma}+H_R$ in these states (multiplied by plane waves
$e^{i\vec k \cdot \vec r}$) yields a $4 \times 4$ Hamiltonian

\begin{widetext}
\begin{equation}
H= \left ( \begin{array}{cccc}
0 & \hbar v_F (k_x -i k_y f_0) & 0 & -i \frac {\lambda ^R} 2 ( 1+ f_0 )+ \Delta _0 f_1 e ^{i \chi} \\
\hbar v_F (k_x +i k_y f_0) & 0 &  i\frac {\lambda ^R} 2 ( 1- f_0 ) + \Delta _0 f_1 e ^{i \chi} & 0 \\
0 &  -i\frac {\lambda ^R} 2 ( 1- f_0) + \Delta _0 f_1 e ^{-i \chi} & 0 & \hbar v_F (k_x -i k_y f_0 )      \\
 i \frac {\lambda ^R} 2 ( 1+ f_0  )+ \Delta _0 f_1 e ^{-i \chi} & 0 & \hbar v_F (k_x +i k_y f_0 ) &0
\end{array} \right ),
\label{Heff_rot}
\end{equation}
\end{widetext}
where
$$f_1 \equiv \int_{-L/2}^{L/2} dx \cos\alpha(x)\sin 2\theta(x).$$

Using  degenerate perturbation
theory, we can search for the positions of the
zero modes in the $k_x,k_y$ plane.  Assuming these to be small,
to quadratic order these are given by the solutions to
\begin{eqnarray}
\hbar ^2  v_F ^2 (k_x^2 -f_0 k_y^2) & =& \Delta_0 ^2 f_1^2 -\frac {\lambda_R^2} 4(1-f_0^2)+\Delta_0 f_1 \lambda ^R f_0 \cos{\chi} \nonumber \\
\hbar ^2 v_F^2 k_x f_0 k_y & = & \Delta _0 f_1 \frac {\lambda _R} 2 \sin{\chi}.
\label{uniform_zeros}
\end{eqnarray}

A first observation is that
when $\lambda _R$=0, the zeros are always on the $k_x$ axis, in contrast to the various
results we found in the last section.
For $\chi=\pi/2$ or $3\pi/2$, where the Zeeman 
field vector is transverse to superlattice axis --
a simple model for N\'eel domain walls, as we discuss in the next subsection --
one finds both $k_x$ and $k_y$ are non-zero when $\lambda_R \ne 0$.
This agrees with our analysis of
a model with piecewise constant $\vec\Delta(x)$ which is non-perturbative in this
parameter, discussed below.

The other interesting case is $\sin\chi$=0, for which $\vec{\Delta}(x)$ rotates longitudinally,
a simple model of Bloch domain walls (again, discussed below.)
Because $\sin{\chi}$=0, $k_x$ or $k_y$ must vanish and the zero modes are on one of the axes.
When $\lambda_R$=0, the zero is on the $k_x$-axis.
For $\cos{\chi}=-1$, one finds that $k_x$ moves towards the origin with
increasing $\lambda_R$.  For $\cos{\chi}=1$, $k_x$ initially moves away from
the origin with increasing $\lambda_R$, but when this parameter is large enough
it also moves towards the origin.
In both cases the Dirac points at $\pm k_x$ ultimately merge at the origin at some
critical value of $\lambda_R$, and
then repel back out along the $k_y$ axis as $\lambda_R$ increases further.

\begin{figure}[htbp]
\includegraphics[width=\columnwidth,clip]{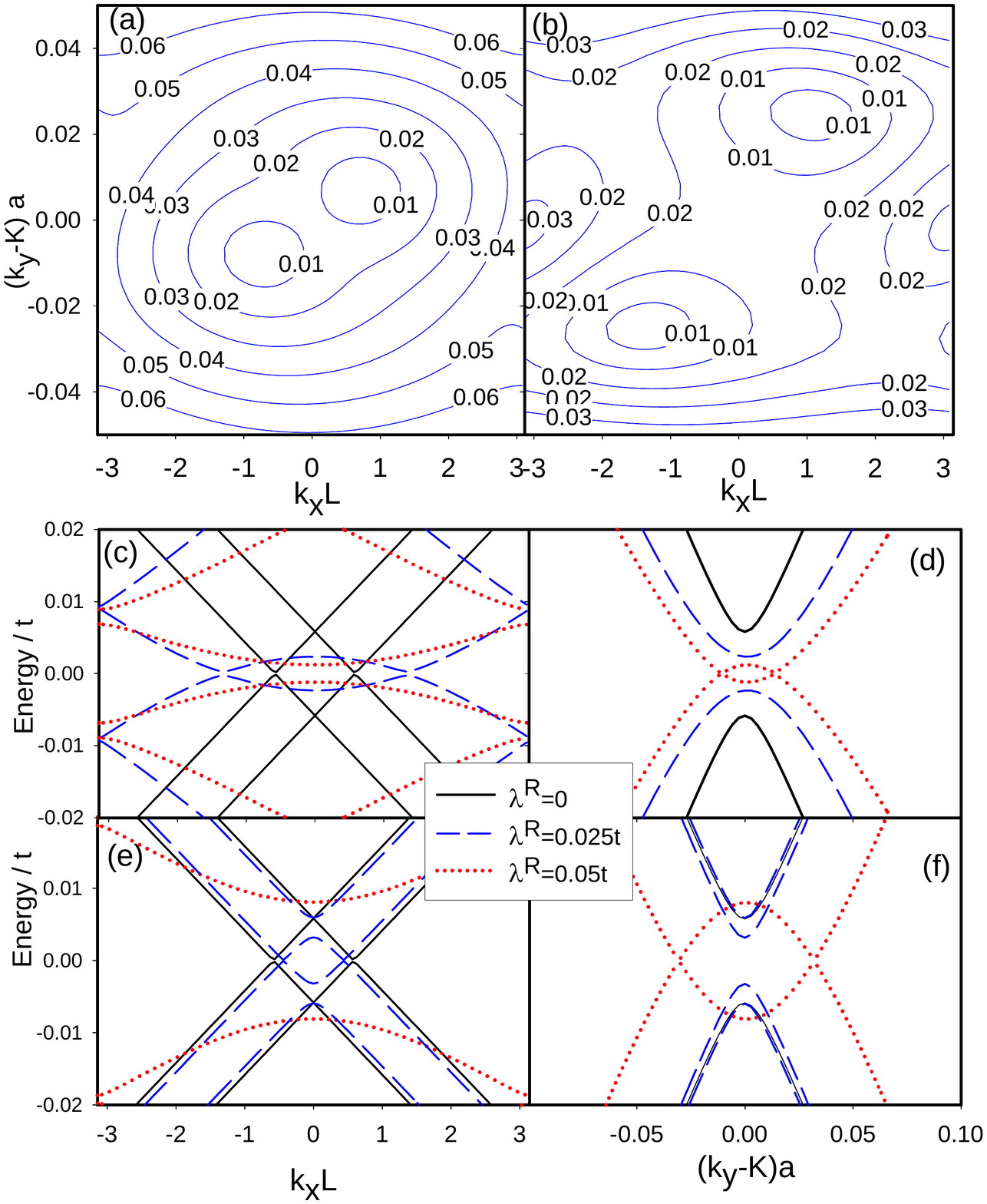}
\caption{(Color online) Numerical tight-binding results near the Dirac point $K$, obtained  with the parameters
$\Delta _0$=0.02$t$ and $L$=86.6$a$, in presence of an uniformly rotating field, Eq. \ref{rotation}. In (a) and (b) we show contour plots of the energy gap for $\chi$=$\pi$/2 and  $\lambda ^R$=0.005t and $\lambda ^R$=0.028$t$ respectively. In this geometry the positions of the Dirac points move in the $k_x$-$k_y$ plane.
In (c)-(d)  and (e)-(f) we plot the band structure in the $k_x$ and $k_y$ directions for $\chi$=0 and $\chi$=$\pi$ respectively, Eq.\ref{rotation}.}
\label{sincos}
\end{figure}

Numerical diagonalization of a tight-binding model with this form of $\vec\Delta(x)$
confirm these expectations from Eq. \ref{uniform_zeros}, indicating
that our perturbative approach yields qualitatively correct results.
Typical results are illustrated in Fig. \ref{sincos}.
We will see in
the next subsection that they are also in agreement with a piecewise constant
$\vec\Delta(x)$ model, for which one may carry out an analysis that is non-perturbative
in all of its components.

\subsection{Piecewise Constant Rotating Exchange Fields: Domain Walls}

In this subsection we consider the impact of in-plane fields when they occur between
regions of constant exchange field $\Delta_z$, which act like Zeeman fields
oriented along the $\hat{z}$ direction.
To describe such situations we generalize this coupling to be formally
the same as for a rotating Zeeman field,
$H_{\Delta} \rightarrow \vec{\Delta}(x)\cdot\vec{\sigma}$ but in
this subsection the rotation is not uniform.
The regions between locations where $\vec{\Delta} \parallel \hat{z}$
are essentially domain walls (DW'S),
which, as mentioned briefly in the previous subsection, may have a variety of forms.
Most prominent are Bloch walls (in which $\vec{\Delta}$ rotates through the
$\Delta_x-\Delta_z$ plane),
and N\'eel walls (in which $\vec{\Delta}$ rotates through the
$\Delta_y-\Delta_z$ plane).  Each
unit cell must contain two DW's, and an additional degree of freedom in this problem
is that the sense of rotation (clockwise or counterclockwise) can be the same or
different.  In the former case, if the DW's have the same
gradient profile in $\vec{\Delta}(x)$,
and the oppositely directed regions of constant $\Delta_z$ have the same magnitude and width,
then the net effective Zeeman field in a unit cell vanishes.
As previewed in Section \ref{sec:uniform_rot} the rotation of the
effective Zeeman field can have a very
interesting impact on where Dirac points reside in the Brillouin zone.
When the DW's
rotate with opposite senses there is a net in-plane effective Zeeman field.
{We will see numerically below that this situation is already interesting
when $\lambda_R=0$: the system when undoped is metallic, supporting overlapping
Fermi surfaces that are electron- and hole-like.}  The inclusion of SOC
however restores the Dirac points, albeit in different locations depending
on whether the $\vec{\Delta}$ rotates through the $\hat{x}$ or $\hat{y}$
direction.

\begin{figure}[htbp]
\includegraphics[width=\hsize,trim = {6cm 8cm 5cm 6cm},clip]{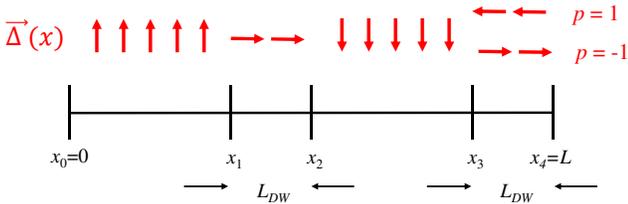}
\caption{(Color online) Illustration of the rotation of $\vec{\Delta}$ in the piecewise
constant $\vec{\Delta}$ model. For $p=1$ the domain walls are oriented in opposite
directions.  For $p=-1$ they are oriented in the same direction.}
\label{DW}
\end{figure}

To examine the electronic spectrum of such structures, we adopt a simple model
in which $\vec\Delta(x)$ is piecewise constant, with four separate regions in
a unit cell: two in which $\vec\Delta \parallel {\hat z}$, and two in which
$\vec\Delta(x)$ is in either the $\hat{x}$ or the $\hat{y}$ direction.  This geometry
is illustrated in Fig. \ref{DW}.  Within the unit cell, $\vec\Delta(x)$ changes
direction at positions $x=x_i$, with $i=0,1,2,3,4$, and the two DW regions have
width $x_2-x_1=x_4-x_3\equiv L_{DW}$.  $\vec\Delta$ in the two DW regions point in the opposite (same)
direction when $p=1$ ($p=-1$). We assume that the two domains
in a unit cell have the same size: $x_1-x_0=x_3-x_2$.

\subsubsection{States and Matching Conditions}

Our strategy will be to construct zero modes for $\lambda^R=0$, and then
add in SOC perturbatively.  We express states of the Hamiltonian in terms of states
which are locally eigenstates of the exchange field.  Writing
$\vec\Delta (x) \equiv |\Delta(x)|\hat{n}(x)$, these are the states which
satisfy
\begin{equation}
\hat{n} \cdot \vec{\sigma} \vec\chi_{\hat n}^{(\pm)}=\pm \vec\chi_{\hat n}^{(\pm)}.
\end{equation}
For $\lambda^R=0$ we anticipate finding zero energy states at $k_y=0$
(although these turn out to be Dirac points only for $p=1$), with wavefunctions
$\vec\psi(x)$ which are annihilated by
\begin{equation}
H'=v_F(-i\tau_x \partial_x) +\vec\Delta(x) \cdot \vec\sigma (x),
\label{Hprime}
\end{equation}
where we have set $\hbar=1$.
It is clear that these zero energy states of $H'$ have $\tau_x$ as a good quantum number.
Writing $\vec\psi(x) = \sum_{s=\pm}u_s(x)\vec\chi_{\hat n}^{(s)}$,
one finds in regions of constant $\hat{n}(x)$ that 
\begin{equation}
u_{\pm}(x) = u_{\pm}(x_0)\exp\left\{\pm i \tau_x\int_{x_0}^x dx' |\vec\Delta(x')| \right\}.
\end{equation}
We next define the set of Pauli matrices $\vec\mu$ such that
$\mu_z\vec\chi_{\hat n}^{(s)} = s \vec\chi_{\hat n}^{(s)}$ for any $x$; in this representation
the quantization axis rotates with $\hat{n}(x).$
Using these, it is not difficult to derive a matching formula across a jump
in the direction of $\hat{n}(x)$ at a point $x_0$ through an angle $\Delta\theta$
in the plane of the initial and final directions of $\hat n(x_0^{\pm}\equiv x_0 + 0^{\pm})$.
Specifically,
\begin{equation}
\vec u(x_0^+) = e^{-{i\over 2} \Delta\theta \mu_y} \vec u(x_0^-),
\label{match}
\end{equation}
where $\vec u \equiv (u_+^*,u_-^*)^{\dag}$.

For the regions $x_j<x<x_i$ in which $n(x)$ is constant, the wavefunctions may be written
\begin{eqnarray}
u_+(x) &=& u_{+,ij}^{(0)}e^{-i\tau_x\theta(x,x_j)} \nonumber \\
u_-(x) &=& u_{-,ij}^{(0)}e^{i\tau_x\theta(x,x_j)},
\end{eqnarray}
where $\theta(x,x_j) = |\vec\Delta|(x-x_j)/v_F$.  Application of the matching condition
Eq. \ref{match} at $x=x_1,\,x_2,\,x_3,\,x_4$ leads to the condition
\begin{equation}
\vec{u}(x=x_4^+)=P_{43}P_{32}P_{21}P_{10} \vec{u}(x=x_0^+) \equiv P \vec{u}(x=x_0^+),
\label{matching_requirement}
\end{equation}
with matrices
\begin{equation}
P_{ij} =
{1 \over {\sqrt{2}}}\left(
\begin{array}{cc}
e^{-i\tau_x\theta(x_i,x_j)} & -e^{i\tau_x\theta(x_i,x_j)} \\
e^{-i\tau_x\theta(x_i,x_j)} & e^{i\tau_x\theta(x_i,x_j)}
\end{array}
\right)
\label{P10P21}
\end{equation}
for $(i,j)=(1,0)$ and $(2,1)$, and
\begin{equation}
P_{ij} =
{1 \over {\sqrt{2}}}\left(
\begin{array}{cc}
e^{-i\tau_x\theta(x_i,x_j)} & -pe^{i\tau_x\theta(x_i,x_j)} \\
pe^{-i\tau_x\theta(x_i,x_j)} & e^{i\tau_x\theta(x_i,x_j)}
\end{array}
\right)
\label{P10P21}
\end{equation}
for $(i,j)=(3,2)$ and $(4,3)$, with $p=1(-1)$ if the DW in $x_3<x<x_4$ rotates
in a clockwise (counterclockwise) sense (see Fig. \ref{DW}).  To obtain a consistent
solution, Bloch's theorem requires that $P$ has an eigenvector with eigenvalue
$-pe^{ik^*_xL}$, where $k^*_x$ must be real.  Using the fact that ${\rm Det}P_{ij}=1$,
it is easy to show that the eigenvalues of $P$ have the form $\lambda_{\pm}=
{1 \over 2} [T \pm (T^2-4)^{1/2}]$, where $T$ is the trace of $P$.  For the
model illustrated in Fig. \ref{DW},
in which $\theta(x_1,x_0)=\theta(x_2,x_1) \equiv \theta_{10}$ and
$\theta(x_2,x_1)=\theta(x_4,x_3) \equiv \theta_{21}$, with some algebra
one finds
\begin{equation}
T=-2\left[\cos 2\theta_{10}\sin^2\theta_{21}+p\cos^2\theta_{21}\right].
\label{T}
\end{equation}
For either sign of $p$ it is easy to see $|T| \le 2$, so that $|\lambda_{\pm}| = 1$
and $k_x^*$ is real.  Thus the system always supports zero modes for $\lambda^R=0$.

\begin{figure}[htbp]
\includegraphics[width=\hsize,trim = {6cm 7cm 4.5cm 6cm},clip]{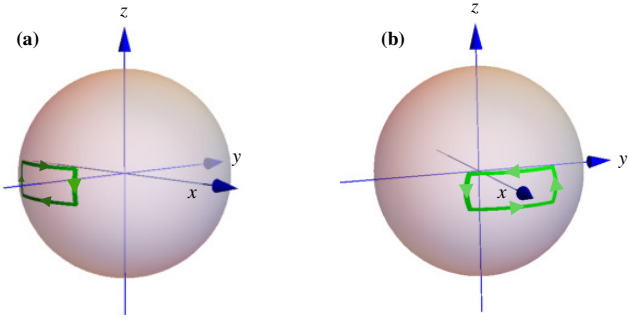}
\caption{(Color online) Effective spinor paths followed by the zero modes of $H'$
(Eq. \ref{Hprime})
when $\vec{\Delta}(x)$ rotates between the $\hat{z}$ and $\hat{x}$ directions (Bloch walls).
In (a) the DW's point in opposite directions ($p=1$); in (b) they point in the same
direction ($p=-1$).}
\label{spherePath}
\end{figure}

\subsubsection{$k_x^*$ and Spinor Precession}

It is interesting that with in-plane effective Zeeman fields, the zero modes end up
at $k^*_x \ne 0$, which was not the case when the Zeeman field was strictly in the
$\hat{z}$ direction.  This behavior has an interesting interpretation.  The zero modes
of $H'$ (Eq. \ref{Hprime}) must solve a first-order differential equation,
$i\partial_x\vec{\psi}={{\tau_x} \over {v_F}} \vec{\Delta(x)}\cdot \vec{\sigma} \vec{\psi}$,
whose solution is formally equivalent to the evolution of a spinor in a time-dependent
magnetic field, if one identifies $x$ as a time coordinate.
In our model of
piecewise-constant $\vec\Delta(x)$, this evolution is a precession of the spinor
around the locally constant magnetic field direction, and because of the boundary
condition, allowed solutions must correspond to orbits that close on themselves.
Fig. \ref{spherePath} qualitatively illustrates such paths for Bloch domain walls,
for the cases $p=1$ (a) and $p=-1$ (b).  In both the trajectories are formed
by alternating rotations around the $\hat{z}$ and $\hat{x}$ directions, but in the
$p=-1$ case the $\hat{x}$ rotations have the same sense while in the $p=1$ case
they have opposite senses.  Because these loops enclose non-vanishing areas,
the wavefunctions pick up a non-trivial phases in going from $x=0$ to $x=L$,
so that the solutions have $k_x^* \ne 0$.  This is in contrast to what must
happen when $\vec\Delta(x)$ is along the $\hat{z}$ direction throughout its
evolution, in which case the trajectory is a single line segment
along which the trajectory rocks back and forth.  This has vanishing enclosed area
so that $k_x^*=0$, consistent with what we found for the zero modes in the last section.
Thus the value of $k_x^*$ at which the zero modes appear are a direct measure
of the non-trivial phase accumulated when a spin degree of freedom traverses an
open loop.

Three comments are in order.  Firstly,
for each of the loops shown in Fig. \ref{spherePath}, there is a second
solution orbiting the same axis but on the other side of the sphere.  These solutions
accumulate the opposite phase of the ones shown, so that one finds solutions at
both $k_x^*$ and $-k_x^*$.  Second, because the solutions 
shown are for fixed $\tau_x = \pm 1$,
for each of $k_x^*$ and $-k_x^*$ there are two degenerate solutions.  
Finally, it must be emphasized that the fact that
these are zero modes does not dictate that they are Dirac points.  We shall
see below that they are for $p=1$, but for $p=-1$, the are individual points
along zero energy contours in the $(k_x,k_y)$ plane.  However, for $\lambda^R>0$ the
latter surfaces become gapped, except for two Dirac points.

\vskip 0.5cm

\subsubsection{Perturbative Treatment of SOC}

SOC admixes the zero energy modes, with different possible effects,
which we evaluate within perturbation theory.
To do so, we write the Hamiltonian in the vicinity of $\vec k = (k_x^*,0) \equiv \vec k^*$ as
\begin{equation}
H_{\vec k^*} = v_F(\vec k^* + \vec q)\cdot\vec\tau + \vec\Delta(x) \cdot \vec\sigma +
\frac{\lambda^R}{2}\left(\sigma_x\tau_y-\sigma_y\tau_x \right).
\label{HDW}
\end{equation}
By construction we have two zero energy states at $\vec k^*$ for $\lambda^R=0$, one
for each value of $\tau_x$, which we write as kets in the form $|\tau_x \rangle$,
where it is implicit that these two states correspond to a specific choice of $\vec k^*$.
Projecting Eq. \ref{HDW} into this subspace, one obtains
\begin{widetext}
\begin{equation}
\bar{H}_{\vec k^*} = \left(
\begin{array}{cc}
v_Fq_x - \frac{\lambda^R}{2} \langle 1 | \sigma_y | 1 \rangle &
    iv_Fq_y\langle 1| \tau_y |-1 \rangle
    + i\frac{\lambda^R}{2} \langle 1 | \sigma_x\tau_y | -1 \rangle \\
-iv_Fq_y\langle -1| \tau_y | 1 \rangle - i\frac{\lambda^R}{2} \langle -1 | \sigma_x\tau_y | 1 \rangle  &
    -v_Fq_x + \frac{\lambda^R}{2} \langle -1 | \sigma_y | -1 \rangle
\end{array}
\right),
\label{projected_H}
\end{equation}
with energy spectrum
\begin{equation}
\varepsilon({\vec q}) = \frac{\lambda^R}{4}\Bigl(- \langle 1 | \sigma_y |1 \rangle
+ \langle -1 | \sigma_y | -1 \rangle \Bigr)
\pm \left\{ \left[ v_Fq_x - \frac{\lambda^R}{4} \langle 1 | \sigma_y |1 \rangle
+ \frac{\lambda^R}{4} \langle -1 | \sigma_y | -1 \rangle \right]^2
+\Bigl|v_Fq_y\langle -1| \tau_y | 1 \rangle
+ \frac{\lambda^R}{2} \langle -1 | \sigma_x\tau_y | 1 \rangle \Bigr|^2 \right\}^{1/2}.
\end{equation}
\end{widetext}
The qualitative effect of SOC on the Dirac points thus depends on just a few matrix elements.
If \hbox{$\langle 1 | \sigma_y |1 \rangle - \langle -1 | \sigma_y | -1 \rangle \ne 0$},
they are shifted away from zero energy.  For
$\langle 1 | \sigma_y |1 \rangle + \langle -1 | \sigma_y | -1 \rangle \ne 0$,
their position along the $k_x$ axis is changed, while
${\rm Re} \left(\langle -1| \sigma_x\tau_y| 1 \rangle/\langle -1| \tau_y| 1 \rangle  \right) \ne 0$ shifts them off the $k_x$ axis
onto a finite $k_y^*$.
Finally
${\rm Im} \left(\langle -1| \sigma_x\tau_y| 1 \rangle/\langle -1| \tau_y| 1 \rangle \right)\ne 0$ opens a gap in the spectrum.

What one needs to know about these various matrix elements can be determined with the help
of Fig. \ref{DW} and by noting that for $\lambda^R=0$ the Hamiltonian has chiral
operators for the N\'eel and Bloch wall cases.  For the former case,
$\chi_{N} \equiv \sigma_x\tau_z$ obeys $\{H',\chi_{N}\}=0$, whereas in the
latter, $\chi_{B} \equiv \sigma_y\tau_z$ obeys $\{H',\chi_{B}\}=0$.  This
means (for the appropriate case) we can relate the basis states by
$| s \rangle = \chi_{N,B} |-s \rangle$, with $s = \pm 1$.  Note these choices of
chiral operators preserve the property $\tau_x| s \rangle = s | s \rangle$.
There are four situations to consider:

\noindent({\it i}) {\it Bloch walls, $p=1$} -- In this case the Bloch walls are
oppositely oriented.  As illustrated in Fig. \ref{spherePath}(a),
the wavefunction describes a trajectory that symmetrically encloses $\hat{y}$ axis
in the spin-space, so for example $\langle 1 | \sigma_y | 1\rangle \ne 0$.
Using $\chi_B | 1\rangle = | -1\rangle$, $[\chi_B,\sigma_y]=0$, and $\chi_B^2=1$,
we see $\sum_s \langle s | \sigma_y | s \rangle \ne 0$, and
$\sum_s s \langle s | \sigma_y | s \rangle = 0$.  We also have
\begin{eqnarray}
\langle -1 | \sigma_x\tau_y | 1 \rangle &=& \langle 1 |\sigma_y\tau_z \sigma_x\tau_y | 1 \rangle \nonumber \\
&=& \langle 1 | \sigma_z\tau_x | 1 \rangle \nonumber \\
&=& \langle 1 | \sigma_z | 1 \rangle = 0. \nonumber
\end{eqnarray}
The last equality follows from the symmetry of the spinor orbit across the $\hat{x}-\hat{y}$
plane in Fig. \ref{spherePath} (a).  The net effect is that the Dirac points are shifted
along the
$k_x$ axis, but the spectrum is qualitatively the same as in the absence of SOC.

\noindent ({\it ii}) {\it Bloch walls, $p=-1$} -- Here the Bloch walls have the same
orientation.  As in the previous case,
$\langle -1 | \sigma_x\tau_y | 1 \rangle =0$, but now the trajectories encircle
the $\hat x$ axis [Fig. \ref{spherePath}(b)], so that $\langle \pm 1 | \sigma_y | \pm 1\rangle = 0$.  The SOC has no qualitative effect on the result and we again find zero energy states
on the $k_x$ axis.  We will verify numerically that these are in fact Dirac points.

\noindent ({\it iii}) {\it N\'eel walls, $p=1$} -- The $\sigma_y$ fields are
oppositely oriented here.
In this case the spin trajectories
surround the $\hat x$ axis, so $\langle s | \sigma_y | s \rangle =0$.  We also
have
\begin{eqnarray}
\langle -1 | \sigma_x\tau_y | 1 \rangle &=& \langle 1 |\sigma_x\tau_z \sigma_x\tau_y | 1 \rangle \nonumber \\
&=& -i\langle 1 | \tau_x | 1 \rangle = -i\nonumber
\end{eqnarray}
and
\begin{eqnarray}
\langle -1 | \tau_y | 1 \rangle &=& \langle 1 |\sigma_x\tau_z\tau_y | 1 \rangle \nonumber \\
&=& -i\langle 1 | \sigma_x | 1 \rangle,\nonumber
\end{eqnarray}
which is also pure imaginary.
In this case the Dirac point is shifted to a non-vanishing value of $k_y^*$.

\noindent ({\it iv}) {\it N\'eel walls, $p=-1$} -- This realization has a
net $\sigma_y$ within a unit cell because the DW's have the same orientation.
The spin trajectories
surround the $\hat y$ axis, and
$\langle s | \sigma_y | s \rangle =-\langle -s | \sigma_y | -s \rangle$.
The results $\langle -1 | \sigma_x\tau_y | 1 \rangle = -i$ and
$\langle -1 | \tau_y | 1 \rangle = -i\langle 1 | \sigma_x | 1 \rangle$
are unchanged
from case ({\it iii}) above; however, here $\langle 1 | \sigma_x | 1 \rangle=0$.
The two eigenvalues of Eq. \ref{projected_H} are then
\begin{equation}
\label{evalsCase4}
\varepsilon(\vec{q})=\frac{\lambda^R}{2}\langle 1 |\sigma_y| 1 \rangle
\pm
\sqrt{\left(\frac{\lambda^R}{2} \right)^2 + v_F^2q_x^2}.
\end{equation}
Along the $k_x$ axis a gap has opened up, but with the two energy eigenvalues
no longer equal and opposite.  However, the opening of a gap at this point
does not imply that the system has become insulating.  We address what actually happens
next using numerical solutions of the tight-binding model.

\subsection{Tight-Binding Results}

\begin{figure}[htbp]
\includegraphics[width=\columnwidth,clip]{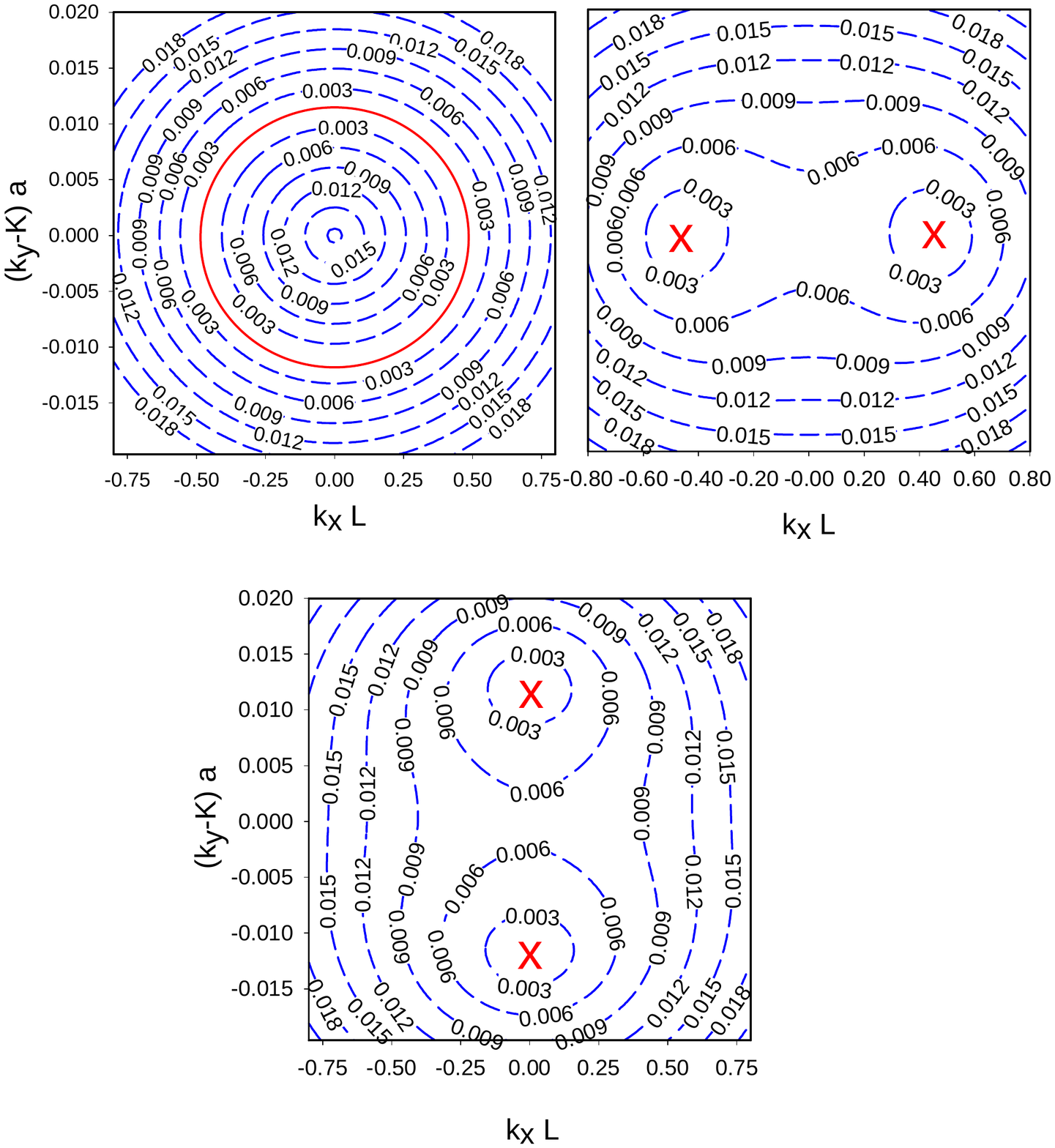}
\caption{(Color online) Contour plots of the energy gap obtained by diagonalizing the tight-binding Hamiltonian  in presence of a piecewise constant rotating Zeeman field with $p$=-1, see Fig \ref{DW}. The calculations use  $\Delta _0$=0.02$t$ and $L$=43.3$a$. In the numerical calculation the four regions appearing in Fig. \ref{DW} all
have the same width. In (a), the Rashba spin orbit coupling is zero and the system is metallic with a Fermi loop marked in red in the contour plot. For finite Rashba coupling the system becomes a semimetal. In  (b) and (c) we show constant energy contour plots  for  Bloch and
N\'eel domain walls respectively, in both cases with $\lambda _R$=0.03$t$ . In (b) and (c) the red crosses indicate the position of the Dirac points.}
\label{DW-KP}
\end{figure}

We have performed tight-binding calculations for the configurations shown in Fig \ref{DW}. 
For domain walls with in-plane fields in opposite
directions ($p$=+1), the results 
are similar to those obtained with an uniformly rotating exchange field (Fig.\ref{sincos}) and are in agreement with the discussion above: for $\lambda^R=0$ one finds Dirac points on
the $k_x$ axis, which for small $\lambda^R$ remain there in the Bloch wall case, and are
shifted onto the $k_x-k_y$ plane in the N\'eel wall case.  For $p=-1$ -- in-plane fields
in the same direction -- one obtains very different results even for $\lambda^R = 0$.
This is illustrated in Fig. \ref{DW-KP}(a), which shows the surface $\epsilon(\vec{k})=0$
forms a closed loop around the $K$ point.  The valence and conduction bands actually
cross along this loop, so that the system is a metal, and all the states on this
constant energy surface are doubly degenerate.  Note that the picture is identical
for both Bloch and N\'eel walls because without SOC, the two structures are related
by spin-rotational symmetry.  

When SOC is added, the two degenerate states along the loops repel,
except at individual points, changing the system into a Dirac semimetal.  For the
Bloch case the resulting Dirac points are on the $k_x$ axis, in agreement with case ($ii$)
above.  This is illustrated in Fig. \ref{DW-KP}(b).  In the
N\'eel case, there are no zero energy states on the $k_x$ axis, and the positive
and negative energy states appear asymmetrically around $\epsilon = 0$,
again in agreement with the analysis above [case ($iv$)].
However, there are nevertheless Dirac points
on the $k_y$ axis, as illustrated in Fig. \ref{DW-KP}(c).
Clearly these are not captured by the perturbative analysis above.

Finally it is interesting to compare and contrast the results for $p=-1$ DW's with those
of uniform Zeeman fields in the $\hat x$ and $\hat y$ directions discussed in Section
\ref{sec:uniform_in_plane} above.  In the latter case (N\'eel walls) the results
are quite similar to simply applying a uniform field in the $\hat y$ direction
coupling to the spin, for which one finds Dirac points on the $k_y$ axis.
In the former (Bloch wall) case, however, whereas the uniform field resulted
in metallic behavior -- zero energy states forming finite size loops in momentum
space -- when the in-plane field is made periodic, the states admix and repel,
except at the two points on the $k_x$ axis, turning the system into a Dirac semimetal.

\section{Summary and Conclusion}

In this study we have demonstrated that graphene with induced spin-orbit coupling
and a periodic Zeeman field supports a remarkably rich set of possible behaviors near
zero energy.  Depending on how the Zeeman field is arranged, one may obtain a metal,
an insulator, or (in most cases)
a Dirac semimetal.  The positions of the Dirac points in this last case may be adjusted 
by varying the relative strengths of SOC ($\lambda^R$)
and the periodic Zeeman field ($\vec{\Delta}(x)$), or the
precise way in which $\vec{\Delta}(x)$ varies in the unit cell.  Unlike the case
of $\lambda^R=0$, the $K$ and $K'$  points are always gapped when there is SOC.

We find that when $\vec{\Delta}(x) \parallel \hat{z}$ everywhere in the unit cell,
the Dirac points are always on the $k_y$ axis, and in most cases there are four
of these, two each on either side of the $K$ and $K'$ points.  While for $\lambda^R = 0$
there are higher order Dirac points along the same direction in momentum space, these
are usually eliminated by SOC, although for narrow ranges of parameters there can be
more Dirac points.  When the possibility of in-plane fields due to a rotating
$\vec{\Delta}(x)$ is allowed, we find the system is generically a Dirac semimetal,
although the positions of the Dirac points in the Brillouin zone are sensitive
to the details of how such rotations are realized.  

Certainly the simplest way to introduce in-plane fields is via a uniform magnetic field
imposed in addition to any effective periodic Zeeman field in the system.  When the
latter is purely in the $\hat z$ direction, we found that the system can be made
into a metal, a Dirac semimetal, or an insulator, by varying only the direction
of the uniform field.  In the last case the band structure of the system carries
non-trivial topology, and we demonstrated that the system supports a quantized
anomalous Hall effect.  

It will be interesting to examine transport in this system, to ascertain
what signatures the changes in the spectrum as $\vec{\Delta}(x)$ is varied may  
present in such experiments.  Another interesting related direction would be
to examine whether such physics occurs on the surfaces of topological insulators, 
or in thin film topological insulators, where
SOC is intrinsic and one need not induce it artificially as in graphene.  Finally,
the question of how imperfectly formed superlattices behave in terms of their
spectra and transport will be particularly to relevant to any experiments
on systems such as what we have studied in this work.  We leave
these as problems for future research.

\section{Appendix: Zero Modes on $k_y$ Axis: $\Delta_z$ Model}
\label{Appendix:Delta_z}

In this appendix, we describe a transfer matrix analysis through which one may
identify zero modes as a function of $k_x,\,k_y$ for this system, and show
specifically that in the absence of in-plane Zeeman fields such modes are
generically present.  The calculation is considerably more involved than the
perturbative analysis described in the main text, but demonstrates that the
basic observation -- the presence of (gapless) Dirac points on the $k_y$ axis
when both a periodic Zeeman field in the $\hat{z}$ direction and Rashba
spin-orbit coupling are present --  is valid to all orders in perturbation theory.

We begin by reiterating the Hamiltonian of the system,
\begin{equation}
H= v_F(p_x\tau_x + p_y\tau_y) + \Delta_z(x)\sigma_z
+ \frac{\lambda^R}{2}(\sigma_x\tau_y -\sigma_y\tau_x),
\end{equation}
where it is assumed that $\Delta_z(-x)=-\Delta(x)$.  For simplicity we consider
piecewise constant forms for $\Delta_z(x)$, and we are interested in
spatially periodic realizations, $\Delta_z(x+L)=\Delta_z(x)$.
To solve for zero modes one needs to find solutions to $H\vec\Phi(x) = 0$
that are Bloch functions, which have the form
$$
\vec\Phi(x)=\vec\psi(x)e^{iQ(x)x+ik_yy}
$$
with $\vec\psi(x)$ and $Q(x)$ piecewise constant.

\subsection{Zero Modes in Regions of Constant $\Delta_z$}

Within the regions of constant $\Delta_z$, $\vec\psi(x)$ satisfies $H_Q\vec\psi =0$, with
$$
H_Q = v_F[Q\tau_x +k_y\tau_y] + \Delta_z\sigma_z
+ \frac{\lambda^R}{2}(\sigma_x\tau_y -\sigma_y\tau_x).
$$
Note that $Q$ may be complex, so that $H_Q$ is not generally Hermitian.
Solving this equation may be simplified by noticing that $H_Q$ has a chiral operator
$\chi_Q$ (i.e., $\{\chi_Q,H_Q \}=0$), given by
$$
\chi_Q=(Q\tau_y-k_y\tau_x)(Q\sigma_y-k_y\sigma_x).
$$
The operator $\chi_Q$ is easily diagonalized and has two eigenvalues,
$$\pm K^2 \equiv \pm (Q^2+k_y^2),$$
and because $H_Q$ anticommutes with $\chi_Q$, one knows that an eigenvector of $\chi_Q$
in the $\pm K^2$ sector must either be mapped to the $\mp K^2$ sector by $H_Q$,
or annihilated
by it.  Eigenvectors of $\chi_Q$ may be written in terms of the eigenvectors of
$Q\mu_y-k_y\mu_x$, where $\mu_{x,y,z}$ are generic Pauli matrices [i.e., acting
either on sublattice ($\vec\tau$) or spin ($\vec\sigma$)], which satisfy
\begin{eqnarray}
\left(Q\mu_y-k_y\mu_x\right)
\frac{1}{\cal N}
\left(
\begin{array}{c}
-iQ-k_y \nonumber \\
\pm K \nonumber
\end{array}
\right)
=
\pm K
\frac{1}{\cal N}
\left(
\begin{array}{c}
-iQ-k_y \nonumber \\
\pm K \nonumber
\end{array}
\right),
\end{eqnarray}
where ${\cal N} = [|K|^2+|iQ+k_y|^2]^{1/2}$.  The kets $|\pm,Q\rangle_{\mu}$ represent
eigenvectors corresponding to eigenvalues $\pm K$.  Eigenvectors of $\chi_Q$ with eigenvalue
$K^2$ are then $|+,Q\rangle_{\tau}|+,Q\rangle_{\sigma}$ and $|-,Q\rangle_{\tau}|-,Q\rangle_{\sigma}$,
and those with eigenvalue $-K^2$ are $|+,Q\rangle_{\tau}|-,Q\rangle_{\sigma}$ and $|-,Q\rangle_{\tau}|+,Q\rangle_{\sigma}$.
Writing $|\pm,Q\rangle_{\tau}|\pm,Q\rangle_{\sigma} \equiv |\pm,\pm;Q\rangle$, one finds
\begin{widetext}
\begin{eqnarray}
H_Q |++:Q\rangle &=& \left(\Delta_z + i\frac{\lambda^R}{2}\right) |+-;Q\rangle
+ \left(iv_FK-i\frac{\lambda^R}{2}\right)|-+;Q\rangle, \nonumber \\
H_Q |--;Q\rangle &=& \left(-iv_FK-i\frac{\lambda^R}{2}\right)|+-;Q\rangle
+ \left(\Delta_z + i\frac{\lambda^R}{2}\right) |-+;Q\rangle, \nonumber \\
H_Q |+-;Q\rangle &=& \left(\Delta_z - i\frac{\lambda^R}{2}\right) |++;Q\rangle
+ \left(iv_FK+i\frac{\lambda^R}{2}\right)|--;Q\rangle, \nonumber \\
H_Q |-+;Q\rangle &=& \left(-iv_FK+i\frac{\lambda^R}{2}\right)|++;Q\rangle
+ \left(\Delta_z - i\frac{\lambda^R}{2}\right) |--;Q\rangle. \nonumber \\
\label{hqchimap}
\end{eqnarray}
\end{widetext}
As expected $H_Q$ maps chiral states across sectors.

We can now search for zero energy
energy states within a particular sector.  For example, writing
$$|\psi_+[Q]\rangle
= a|++;Q\rangle +b|--;Q\rangle,$$ using Eqs. \ref{hqchimap} one finds $H_Q$ annihilates
this state if
\begin{eqnarray}
\left(
\begin{array}{c c}
\Delta_z + i\frac{\lambda^R}{2} & -iv_FK-i\frac{\lambda^R}{2} \notag \\
iv_FK-i\frac{\lambda^R}{2} & \Delta_z + i\frac{\lambda^R}{2} \notag \\
\end{array}
\right)
\left(
\begin{array}{c}
a \notag \\
b \notag \\
\end{array}
\right)
=0.\notag \\
\label{matrix_zero}
\end{eqnarray}
Non-trivial solutions of Eq. \ref{matrix_zero} exist when
$
(v_FK)^2=\Delta_z^2+i\lambda^R\Delta_z,
$
or, alternatively,
$$v_FQ=\pm \left[-v_F^2k_y^2+\Delta_z^2+i\lambda^R\Delta_z \right]^{1/2}\equiv \pm v_F\tilde Q.$$
With this technique, we can generate four zero energy states in a region of constant $\Delta_z$:
\begin{eqnarray}
&|\psi_+[\tilde Q]\rangle  \equiv & a|++;\tilde Q \rangle  + b|--;\tilde Q \rangle, \notag \\
&|\psi_+[-\tilde Q]\rangle  \equiv & a|++;-\tilde Q \rangle  + b|--;-\tilde Q \rangle, \notag \\
&|\psi_-[\tilde Q^*]\rangle  \equiv & a^*|-+;\tilde Q^* \rangle  + b^*|+-;\tilde Q^* \rangle, \notag \\
&|\psi_-[-\tilde Q^*]\rangle  \equiv & a^*|-+;-\tilde Q^* \rangle  + b^*|+-;-\tilde Q^* \rangle, \notag \\
\label{delta_const_modes}
\end{eqnarray}
with $a=iv_FK + i\lambda^R/2$, $b=\Delta_z + i\lambda^R/2$, and
$v_FK = \sqrt{\Delta_z^2 + i\lambda^R\Delta_z}$.

\subsection{Generalized Mirror Symmetry}

We next need to match the solutions in Eqs. \ref{delta_const_modes} corresponding to $\Delta_z>0$,
which we can assume for concreteness to be in the region $0<x<L/2$, to those corresponding to
$\Delta_z<0$ in the interval $-L/2<x<0$.  To do so we define $x$-dependent wavefunctions via
$\vec\Psi = \vec\psi(x) e^{iQ(x)x}$, which for zero modes obeys the equation
$H_{k_y}\vec\Psi =0$, with
$$
H_{k_y} = v_F[(-i\partial_x)\tau_x +k_y\tau_y] + \Delta_z(x)\sigma_z
+ \frac{\lambda^R}{2}(\sigma_x\tau_y -\sigma_y\tau_x).
$$
The matching is simplified by noting that, due to the antisymmetry of $\Delta_z(x)$,
$H_{k_y}$ commutes with the generalized mirror operation $X \equiv \tau_y\sigma_xI_x$,
where $I_x$ carries out the spatial mirror inversion, $I_xf(x) \equiv f(-x)$ for any $f(x)$.
Since $X^2=1$, this means we can classify the zero modes into two groups, satisfying
$X\vec\Psi_{\pm} = \pm \vec\Psi$.  The operation is particularly interesting at $x=0$, where
$X\vec\Psi_{\pm}(x=0)=\tau_y\sigma_x\vec\Psi_{\pm}(x=0) \equiv {\cal M}_y\vec\Psi_{\pm}(x=0)$ , so that
at the origin $\vec\Psi$ is an eigenstate of the (purely matrix) operation ${\cal M}_y$.  In terms
of eigenstates of $\tau_z$ and $\sigma_z$, $|s_1,s_2\rangle_0$, where
$\tau_z|s_1,s_2\rangle_0=s_1|s_1,s_2\rangle_0$ and $\sigma_z|s_1,s_2\rangle_0=s_2|s_1,s_2\rangle_0$,
with $s_1,\,s_2 = \pm 1$, the action of ${\cal M}_y$ is
\begin{eqnarray}
\begin{array}{c c}
{\cal M}_y|1,1\rangle_0=i|-1,-1\rangle_0, & {\cal M}_y|-1,-1\rangle_0=-i|1,1\rangle_0, \notag \\
{\cal M}_y|1,-1\rangle_0=i|-1,1\rangle_0, & {\cal M}_y|-1,1\rangle_0=-i|1,-1\rangle_0, \notag \\
\end{array}
\label{my_action}
\end{eqnarray}
which is the action of a Pauli $\sigma_y$ matrix in each of the two-dimensional
sectors defined by $s_1s_2=1$ and $s_1s_2=-1$.  If one orders the basis states as
$$\left(|1,1\rangle, |1,-1\rangle, |-1,1\rangle, |-1,-1\rangle \right),$$
the coefficients for the
states corresponding to those in Eq. \ref{delta_const_modes} are
\begin{widetext}
\begin{eqnarray}
\vec\psi_+[\pm\tilde Q] =
\left(
\begin{array}{c}
(a+b)(\mp i\tilde{Q}-k_y)^2 \nonumber \\
(a-b)(\mp i\tilde{Q}-k_y)K  \nonumber \\
(a-b)(\mp i\tilde{Q}-k_y)K  \nonumber \\
(a+b)K^2 \nonumber \\
\end{array}
\right),
\,\,
\vec\psi_-[\pm \tilde Q^*] =
\left(
\begin{array}{c}
(a^*+b^*)(\mp i\tilde{Q}^*-k_y)^2 \notag \\
(a^*-b^*)(\mp i\tilde{Q}^*-k_y)K^*  \notag \\
-(a^*-b^*)(\mp i\tilde{Q}^*-k_y)K^*  \notag \\
-(a^*+b^*)K^{*2} \notag \\
\end{array}
\right).\\
\label{Delta_const_modes_my}
\end{eqnarray}
\end{widetext}
With some tedious (albeit straightforward) algebra, we can construct from these
eigenstates of ${\cal M}_y$ with eigenvalues $m_y = \pm 1$.  For $m_y=1$ one finds
\begin{widetext}
\begin{eqnarray}
\vec\psi_1[m_y=1]=
-\left(\frac{1-i}{\sqrt{2}}
\right)
\left(
\frac{i\tilde Q-k_y}{4i\tilde QK^2(a+b)}\right)
\vec\psi_+[\tilde Q]
+\left(\frac{1-i}{\sqrt{2}}
\right)
\left(
\frac{-i\tilde Q-k_y}{4i\tilde QK^2(a+b)}\right)
\vec\psi_+[-\tilde Q] \notag\\
-\left(\frac{1+i}{\sqrt{2}}
\right)
\left(
\frac{i\tilde Q^*-k_y}{4i\tilde Q^*K^{*2}(a^*+b^*)}\right)
\vec\psi_-[\tilde Q^*]
+\left(\frac{1+i}{\sqrt{2}}
\right)
\left(
\frac{-i\tilde Q^*-k_y}{4i\tilde Q^*K^{*2}(a^*+b^*)}\right)
\vec\psi_-[-\tilde Q^*],
\label{psi_1}
\end{eqnarray}
\begin{eqnarray}
\vec{\psi}_2[m_y=1]&=&
\frac{1}{\sqrt{2}}
\left[
\left(
\frac{1+i}{2}
\right)
\frac{1}{2(a-b)K^3}
+
\left(
\frac{-1+i}{2}
\right)
\frac{2k_y(a^*+b^*)}{4i\tilde{Q}(a+b)(a^*-b^*)K^{*}K^2}
\right]
\left(
i\tilde{Q}-k_y
\right)
\vec\psi_+[\tilde Q] \notag\\
&+&
\frac{1}{\sqrt{2}}
\left[
\left(
\frac{1+i}{2}
\right)
\frac{1}{2(a-b)K^3}
-
\left(
\frac{-1+i}{2}
\right)
\frac{2k_y(a^*+b^*)}{4i\tilde{Q}(a+b)(a^*-b^*)K^{*}K^2}
\right]
\left(
-i\tilde{Q}-k_y
\right)
\vec\psi_+[-\tilde Q] \notag\\
&+&
\frac{1}{\sqrt{2}}
\left[
\left(
\frac{1+i}{2}
\right)
\frac{-2k_y(a+b)}{4i\tilde{Q}^*(a^*+b^*)(a-b)KK^{*2}}
-
\left(
\frac{-1+i}{2}
\right)
\frac{1}{2(a^*-b^*)K^{*3}}
\right]
\left(
i\tilde{Q}^*-k_y
\right)
\vec\psi_-[\tilde Q^*] \notag\\
&+&
\frac{1}{\sqrt{2}}
\left[
-\left(
\frac{1+i}{2}
\right)
\frac{-2k_y(a+b)}{4i\tilde{Q}^*(a^*+b^*)(a-b)KK^{*2}}
-
\left(
\frac{-1+i}{2}
\right)
\frac{1}{2(a^*-b^*)K^{*3}}
\right]
\left(
-i\tilde{Q}^*-k_y
\right)
\vec\psi_-[-\tilde Q^*]. \notag\\
\label{psi_2}
\end{eqnarray}
\end{widetext}
States with $m_y=-1$ may be obtained from this by defining an operator ${\cal M}_x$ such that
${\cal M}_x|s_1,s_2\rangle=|-s_1,-s_2\rangle$, which anticommutes with ${\cal M}_y$, so that
$|\psi_i[m_y=-1]\rangle \equiv {\cal M}_x|\psi_i[m_y=1]\rangle$.

\subsection{Wavefunctions for $x \ne 0$ and Wavefunction Matching}

In this subsection we describe how one finds values of $k_y$ for which appropriately
continuous wavefunctions with zero energy can be constructed.  In particular we
do so for $m_y=1$;
The zero modes for $m_y=-1$ can be constructed from these, and in particular
will exist at the same values of $k_y$, as we explain momentarily.
Eqs. \ref{psi_1} and \ref{psi_2} represent explicit wavefunctions at $x=0$,
which can
be extended into $x>0$ simply by multiplying each term by the appropriate plane wave.
Defining coefficients $A_i$ by writing
\begin{widetext}
\begin{equation}
\vec{\psi}_i[m_y=1] \equiv A_i[\tilde Q]\vec\psi_+[\tilde Q]
+A_i[-\tilde Q]\vec\psi_+[-\tilde Q]
+A_i^*[-\tilde Q]\vec\psi_-[\tilde Q^*]
+A_i^*[\tilde Q]\vec\psi_-[-\tilde Q^*],
\label{alpha_def}
\end{equation}
one obtains
\begin{equation}
\vec\Psi_i[m_y=1;x] =
A_i[\tilde Q]e^{iQx} \vec\psi_+[\tilde Q]
+A_i[-\tilde Q]e^{-iQx} \vec\psi_+[-\tilde Q]
+A_i^*[-\tilde Q]e^{iQ^*x} \vec\psi_-[\tilde Q^*]
+A_i^*[\tilde Q]e^{-iQ^*x} \vec\psi_-[-\tilde Q^*].
\end{equation}
\end{widetext}
For $x<0$ the wavefunction is obtained using
\begin{eqnarray}
|\Psi_i[m_y;-x]\rangle&=&I_x|\Psi_i[m_y;x]\rangle \notag\\
&=&{\cal M}_yM|\Psi_i[m_y;x]\rangle \notag\\
&=&m_y{\cal M}_y|\Psi_i[m_y;x]\rangle. \notag
\end{eqnarray}
Note that for general values of $x$,
$|\Psi_i[m_y;x]\rangle$ is {\it not} an eigenstate of the ${\cal M}_y$ operator.
However, at two points it is:
$x=0$, where the wavefunction must be continuous [$I_x\vec{\Psi}(x=0^+)=\vec{\Psi}(x=0^-)]$,
and at $x=\pm L/2$, where
$I_x\vec{\Psi}(x=L/2)=\vec{\Psi}(x=-L/2)=e^{ik_xL}\vec{\Psi}(x=L/2)$ due to
Bloch's theorem.  For the present purpose we focus on states with $k_x=0$, anticipating
from our numerical investigations that Dirac points if present are on the $k_y$
axis when there are no in-plane Zeeman fields.  We thus require that eigenstates of the
Hamiltonian which are properly continuous are also eigenstates of ${\cal M}_y$ at $x=L/2$:
\begin{equation}
{\cal M}_y|\Psi[m_y=1;L/2]\rangle=|\Psi[m_y=1;L/2]\rangle,
\label{matching}
\end{equation}
where
\begin{widetext}
$$|\Psi[m_y=1;L/2]\rangle=\sum_{i=1,2} u_i |\Psi_i[m_y=1;L/2]\rangle \equiv
\sum_{i=1,2} u_i \psi_{i}[s_1,s_2]|s_1,s_2\rangle_0$$
for some coefficients $u_i$.  Note the simplification that different values
of $m_y$ are not admixed by the matching conditions.
\end{widetext}
Equating the coefficients of the various $|s_1,s_2\rangle_0$ states on either side of
Eq. \ref{matching} generates four equations, although one quickly recognizes that only two
of these are linearly independent.  Eq. \ref{matching} can thus be satisfied if we can
find coefficients $u_1,\,u_2$ such that
\begin{widetext}
\begin{eqnarray}
\left(
\begin{array}{c c}
\psi_{1}[1,1]+i\psi_{1}[-1,-1] &  \psi_{2}[1,1]+i\psi_{2}[-1,-1] \notag \\
\psi_{1}[1,-1]+i\psi_{1}[1,-1] &  \psi_{2}[1,-1]+i\psi_{2}[-1,1] \notag \\
\end{array}
\right)
\left(
\begin{array}{c}
u_1 \notag\\
u_2 \notag\\
\end{array}
\right)
\equiv
{\bf D}
\left(
\begin{array}{c}
u_1 \notag\\
u_2 \notag\\
\end{array}
\right)
=0. \\
\label{psi_matrix}
\end{eqnarray}
\end{widetext}
The coefficients $\psi_i[s_1,s_2]$ may be obtained explicitly using
using Eqs. \ref{psi_1}, \ref{psi_2}, and \ref{alpha_def}; the expressions are
lengthy and not particularly illuminating, and so
are not provided here.  We note, however, that their forms turn out to
guarantee that the matrix ${\bf D}$ appearing in Eq. \ref{psi_matrix} is purely real.  Finally,
non-trivial solutions to Eq. \ref{psi_matrix} can be found when
$$
D(k_y) \equiv det {\bf D} =0.
$$
The determinant $D(k_y)$ is plotted for various choices of parameters in the main text.

Finally, we come back to the fact that our construction was carried out specifically
for $m_y=1$.  Defining $T_{L/2}$ as a translation operator by half a unit cell
($T_{L/2}f(x)=f(x+L/2)$), it is not difficult to show the $[T_{L/2},I_x]=0$
within the subspace of functions that are periodic ($f(x+L)=f(x)$).  So within
our restriction to Bloch states with $k_x=0$, we can consider the action of the
operator $S \equiv {\cal K}T_{L/2}$, where ${\cal K}$ denotes complex conjugation.
Given a state $\vec\Psi[m_y;x]$ for which $H_{k_y}\vec\Psi[m_y;x] = 0$,
it is easy to show
that $H_{k_y}S\vec\Psi[m_y;x] = 0$.  Moreover,
$XS\vec\Psi[m_y;x] = -m_yS\vec\Psi[m_y;x]$, so $S\vec\Psi[m_y;x]$ lies in
the opposite subspace from $\vec\Psi[m_y;x]$ under $X$ .  Thus we find zero energy
states occurring in pairs with different $X$ eigenvalues, and understand
that the Hamiltonian does not cause level repulsion between them because
$[H_{k_y},X]=0$.

\begin{acknowledgments} This work has been supported by MEyC-Spain under grant FIS2015-64654-P, by Brazilian funding agency Capes,
by the NSF through Grant Nos. DMR-1506263 and
DMR-1506460, and by the US-Israel Binational Science
Foundation.

\end{acknowledgments}

%


\begin{thebibliography}{41}
\bibitem [{not()}]{note1}%
  \BibitemOpen
  \href@noop {} {}\bibinfo {note} {We assume that the magnitude and period of
  the exchage field are weak enough so that no new Dirac points are induced in
  the band structure.}\BibitemShut {Stop}%
\end{thebibliography}
\end{document}